\documentclass[11pt]{article}
\usepackage[margin=1in]{geometry}
\usepackage{amsmath,amsfonts,amssymb}
\usepackage{array}
\usepackage{textcomp}
\usepackage{hyperref}
\usepackage{booktabs}
\usepackage{stfloats}
\usepackage{url}
\usepackage{verbatim}
\usepackage{graphicx}
\usepackage{enumitem}
\usepackage{cite}
\usepackage{formatting/style_cdc}
\usepackage{placeins}
\newcommand{\x}{{\bm x}}
\newcommand{\bu}{{\bm u}}

\usepackage{csvsimple}
\usepackage[utf8]{inputenc}
\usepackage[margin=1in]{geometry}
\usepackage[numbers]{natbib}
\usepackage{floatpag}
\usepackage{float}
\usepackage{wrapfig}

\begin{document}

\title{Empowering Federated Learning with Implicit Gossiping: Mitigating Connection  Unreliability Amidst Unknown and Arbitrary Dynamics }

\author{Ming Xiang \thanks{
A preliminary version of the paper \cite{xiang2023towards} was presented at the IEEE 62nd Conference on Decision and Control 2023, Singapore.
Ming Xiang, Stratis Ioannidis, Edmund Yeh, and Lili Su are with the Department of ECE, Northeastern University, Boston, MA 02115 USA (email: {\tt \{xiang.mi,l.su\}@northeastern.edu; \{ioannidis,eyeh\}@ece.neu.edu}).}
\and
Stratis Ioannidis \footnotemark[1] \and
Edmund Yeh \footnotemark[1] \and
Carlee Joe-Wong \thanks{
Carlee Joe-Wong is with the Department of ECE, Carnegie Mellon University, Pittsburgh, PA 15213 USA (email: {\tt cjoewong@andrew.cmu.edu}).
We gratefully acknowledge the support 
from the National Science Foundation (NSF) under grants 2106891, 2107062,
from the NSF CAREER award under grant 2340482,
and from ARO under contract W911NF-23-2-0014.}  \and
Lili Su \footnotemark[1]
}

\date{March, 2024}
\maketitle

\begin{abstract}
Federated learning %
is a popular distributed learning approach
for training a machine learning model without disclosing raw data. 
It consists of a parameter server and a possibly large collection of clients (\eg, in cross-device federated learning) that may operate in congested and changing environments.  
In this paper, we study federated learning 
in the presence of stochastic and dynamic communication failures wherein the uplink between the parameter server and client $i$ is on 
with {\em unknown} probability $p_i^t$ in round $t$. 
Furthermore, we allow the %
dynamics of $p_i^t$ %
to be {\em arbitrary}. %

We first demonstrate that when the $p_i^t$'s vary across clients, the most widely adopted federated learning algorithm, Federated Average (FedAvg), experiences significant bias. To address this observation, we propose Federated Postponed Broadcast (FedPBC), a simple variant of FedAvg. FedPBC differs from FedAvg in that the parameter server postpones broadcasting the global model till the end of each round. 
Despite uplink failures, we show that FedPBC converges to a stationary point of the original non-convex objective.
On the technical front, postponing the global model broadcasts enables implicit gossiping among the clients with active links in round $t$. Despite the time-varying nature of $p_i^t$, we can bound the perturbation of the global model dynamics using techniques to control gossip-type information mixing errors.
Extensive experiments have been conducted on 
real-world datasets over diversified unreliable uplink patterns to corroborate our analysis.

\end{abstract}

\section{Introduction}
\label{sec: intro}
Federated learning is a distributed machine learning paradigm wherein a parameter server and a collection of end/edge devices (referred to as {\em clients}) collaboratively train a machine learning model without requiring clients to disclose their local data  \cite{mcmahan2017communication,kairouz2021advances}. 
Instead of uploading raw data to the parameter server, the clients work at the front line in processing their local data and periodically report their updates to the parameter server, which then effectively aggregates those 
updates to obtain a new model.
The massive system scale and the client heterogeneity in hardware, software, and environments %
leads to either active \cite{mcmahan2017communication,kairouz2021advances} or passive \cite{Li2020,philippenko2020bidirectional,wang2022} partial client participation,  i.e., in each round, the parameter server receives updates from a subset of clients only.    

\begin{figure}[!tb]
\centering
\includegraphics[width=\columnwidth,trim=6cm 6cm 1cm 9cm,clip]{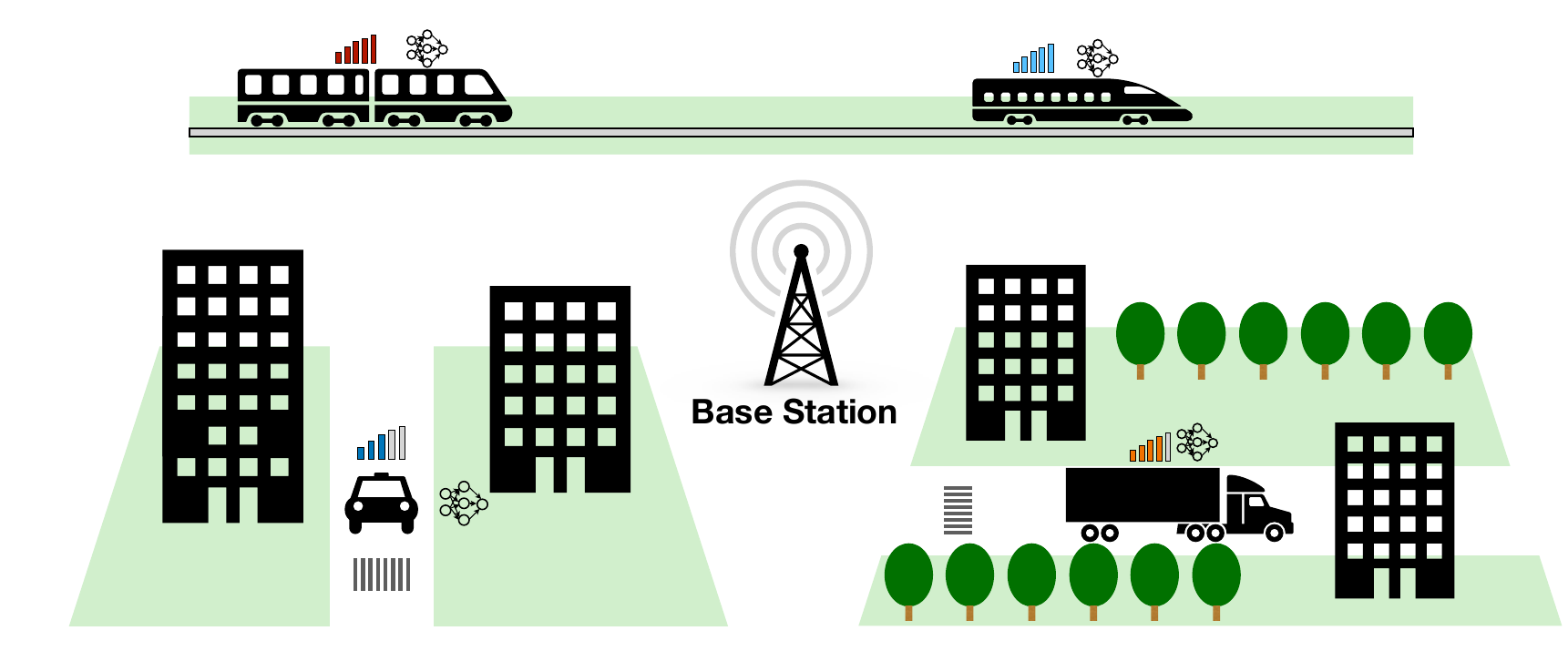}        

\caption{\small
A federated learning system with moving autonomous vehicles as clients.
The signal strength of the vehicles indicates the communication conditions. 
}
\label{fig: moving vehicles in FL}
\label{fig:FL systems}
\end{figure}

Federated learning systems are often deployed in congested and uncontrollable environments with mobile clients %
such as smartphones and other internet-of-thing devices.  
Client mobility and environment complexity can result in unreliable communication \cite{bonawitz2019towards,Ye2022JSTSP,kairouz2021advances}, 
which may even vary significantly
across time and devices.    
For example, the network connection between a smartphone and a base station may be lost when the smartphone is on a train passing through a tunnel. 
Popular transportation layer protocols either have an expensive overhead (such as TCP) or are unreliable (such as UDP) \cite{Ye2022JSTSP}.   
Previous research has demonstrated that unpredictable fluctuations in both the speed and direction of mobile end devices can lead to erratic capacity patterns in 5G links    
\cite{zhang2019will,guan2019effects,mateo2019analysis}.

Unreliable communication in federated learning systems has not caught attention until recently.  
Ye et al.\,\cite{Ye2022JSTSP} assume the communication failures are symmetric with fixed underlying statistics. 
Time-varying communication constraints are considered in \cite{ribero2022federated}, 
wherein the evolution of the feasible client sets is assumed to follow a homogeneous Markov chain with a steady-state distribution. 
Yet, as we shall see from the example illustrated in~\prettyref{fig: moving vehicles in FL}, the assumption of time-invariant communication dynamics easily breaks down when clients are mobile and operate in complex environments. 
More detailed discussions are reserved in~\prettyref{sec: related work}.
It is tempting to address dynamic communication capabilities via asynchronous distributed learning,  wherein an active client contributes to the global model only
when its uplink is on. 
Unfortunately, 
to the best of our knowledge, 
existing literature mostly assumes
bounded delay assumption of the uplink availability \cite{agarwal2011distributed,lian2015asynchronous,zhang2020taming,feyzmahdavian2016asynchronous,arjevani2020tight,stich2020error,yang2022anarchic},
which are hard to hold 
in practical federated learning systems \cite{gu2021fast,kairouz2021advances}. 
Often,
clients in a federated learning system communicate with the parameter server
on their own schedule, 
which
is subject to communication constraints and
can have variations due to hardware or software heterogeneity.

In this paper, we study stochastic %
uplink failures wherein the uplink between the parameter server and client $i$ is active with probability $p_i^t$ in %
round $t$.  
Furthermore, we allow $p_i^t$ to be time-varying and its dynamics to be {\em unknown} and {\em arbitrary}. %
An illustrative example that motivates our problem formulation is shown in~\prettyref{fig: moving vehicles in FL}. 
Specifically, 
fast-moving vehicles quickly pass through a base station's coverage, resulting in frequent handovers.
Varying road conditions (\eg, tall buildings, tunnels), traffic densities, and unforeseeable extreme weather can lead to complex dynamics of the connection probabilities.    
To the best of our knowledge, understanding the convergence of federated learning in the presence of such stochastic uplink failures remains largely under-explored. 

\vskip 0.5\baselineskip
\noindent{\bf Contributions.} 
Our contributions are three-fold: 
\begin{itemize}[leftmargin=*]
\item We identify simple instances and show both analytically and numerically that when the $p_i^t$'s are not uniform, {\em Federated Average} (FedAvg) -- the most widely adopted federated learning algorithm -- fails to minimize the global objective even for simple, convex loss function.
\item 
We propose {\em Federated Postponed Broadcast} (FedPBC),  which %
differs from FedAvg in that the parameter server postpones broadcasting the global model till the end of each round. 
We show in~\prettyref{thm: main} that, in expectation, FedPBC converges to a stationary point of the non-convex global objective. %
The correctness of our FedPBC 
{\em neither} imposes any ``balancedness'' requirement on $p_i^t$'s {\em nor} requires the stochastic gradients or their noises to be bounded. 
On the technical front, %
postponing the global model broadcasts enables implicit gossiping among the clients with active links. 
Hence, %
the perturbation %
caused by non-uniform and time-varying $p_i^t$ can be bounded by leveraging the techniques of controlling information mixing errors. 
It is worth noting that as long as $p_i^t\ge c$ for an absolute constant $c$, 
the staleness of uplink availability is upper bounded (see~\prettyref{prop: staleness term}). 

\item 
We validate our analysis empirically
on three real-world datasets.
Extensive experiments are conducted 
on both {\em time-varying} and {\em time-invariant} Bernoulli, Markovian, and cyclic uplink unreliable patterns. %
\end{itemize}

\section{Related Work}
\label{sec: related work}
In this section, we explore additional related work and present an exhaustive discussion on relevant work mentioned in Section \ref{sec: intro}. 
The section is divided into two parts:
client unavailability and bias correction in distributed learning.
\subsection{Client Unavailability}
The communication unreliability addressed in this paper is implicitly linked to client unavailability. 
The key commonality is that, during failure occurrences, the parameter server cannot receive responses from the involved clients.
Prior literature can roughly be categorized into two groups:  {\em known client participation statistics} %
\cite{mcmahan2017communication,Li2020,perazzone2022communication,cho2022towards,pmlr-v202-cho23b,chen2022optimal,ruan2021towards}
and {\em unknown client participation statistics} %
\cite{gu2021fast,yan2023federated,wang2022,ribero2022federated,wang2023lightweight}.

\vspace{0.5em} 

\noindent{\bf Known client participation statistics.} 
In the seminal works of federated learning \cite{mcmahan2017communication,Li2020}, the parameter server proactively 
determines ``who to participate'' via sampling the clients 
either uniformly at random or proportionally to clients' local data volume.  
A more challenging yet practical scenario where the parameter server loses such proactive selection capability is considered in~\cite{Li2020,philippenko2020bidirectional,kairouz2021advances,pmlr-v180-jhunjhunwala22a}.  
To limit the negative impacts of stragglers, 
the parameter server 
only waits for a few fastest client responses before moving to the next round.  
To balance the contribution of active and inactive clients, 
the parameter server adjusts their aggregation weights according to the corresponding response probabilities, 
which are assumed to be known. 
On the other hand,
some research aims to {\em manipulate} client scheduling schemes to either improve communication efficiency or to speed up training, 
where, at a high level, 
clients are required to participate
whenever the parameter server requests.
In contrast, 
clients are allowed to communicate on their own schedules in our work.
To name a few,
Perazzone et al.~\cite{perazzone2022communication} analyze the convergence of FedAvg under time-varying client participation rates. Nevertheless, they assume
(1) the participation rates $p_i^t$'s are a known prior %
and 
(2) the parameter server controls the participation rates to save communication bandwidth. %
Chen et al.~\cite{chen2022optimal} study a client sampling scheme under which 
the parameter server only samples the most important updates. 
Toward this, 
the parameter server needs to calculate and manipulate the participation rates. 
Cho et al.~\cite{cho2022towards} devise an adaptive client sampling scheme that non-uniformly selects active clients in each round to accelerate training. 
Unfortunately, the convergence is up to a non-vanishing error.
In another work, Cho et al.~\cite{pmlr-v202-cho23b} study a cyclic participation scheme to accelerate FedAvg training, where the parameter server designs and controls the cyclic participation pattern of the clients. 
Tang et al.~\cite{tang2023tackling} utilize the notion of system-induced bias, where the local data set of active clients does not represent the entire population due to time-varying unbalanced communications.
Albeit facing similar time-varying communications, 
their approach requires, which we do not, the parameter server to select the representative clients strategically.

\vspace{0.5em} 
\noindent{\bf Unknown client participation statistics.}
Only a handful of existing works fall under this line of research.
Wang et al. \cite{wang2022} consider structured client unavailability.  
For the methods in \cite{wang2022} to converge to stationary points, 
the response rates of the clients need to be ``balanced'' in the sense that 
either (1) the $p_i^t$'s are deterministic and satisfy the regularized participation, i.e., there exists $\mu>0$ such that $\frac{1}{P}\sum_{\tau=1}^P p_i^{t_0+\tau} =  \mu$ for all clients at all $t_0\in \{0, P, 2P, \cdots\}$ 
where $P$ is some carefully chosen integer; or (2) $p_i^t$'s are random and satisfy $\expect{p_i^t} = \mu$ for all clients and sufficiently many rounds.
In contrast, we do not require such probabilistic ''balanceness''.
Ribero et al. \cite{ribero2022federated} consider random client availability whose underlying response rates are also heterogeneous and time-varying with unknown dynamics. 
The key difference from our focus is that the underlying dynamics of $p_i^t$ in \cite{ribero2022federated} is assumed to be Markovian with a unique stationary distribution,
which is hard to justify when the dynamics vary significantly.  
Gu et al.\,\cite{gu2021fast} consider general client unavailability patterns for both strongly convex and non-convex global objectives. 
For non-convex objectives (which is our focus), they require that the consecutive unavailability rounds of a client to be deterministically upper bounded, which does not hold even for the simple uniform and time-invariant response rates. 
Moreover, they require the noise of the stochastic gradient to be uniformly upper-bounded. %
Wang et al.\,\cite{wang2023lightweight} design a lightweight algorithm to fix FedAvg over non-uniform participation probabilities.
However, their analysis is applicable only to time-invariant communications. 
\subsection{Bias Correction in Distributed Learning}
As we will show in \prettyref{sec: case study counterexample}, 
FedAvg suffers significant bias when the uplinks are non-uniformly available.
However, the term bias is not new and has different meanings under different contexts in the field of distributed learning.
For example, clients perform multiple local updates to save communication in federated learning before communicating with the parameter server.
Yet, bias arises when clients are heterogeneous in the number of local steps~\cite{wang2020tackling}. 
To correct the bias, Wang et al.~\cite{wang2020tackling} %
propose FedNova \cite{wang2020tackling}, in which every client participates, and the parameter server normalizes the contribution of different clients by adjusting the aggregation weights according to their numbers of local steps.   
In fully distributed %
settings (where no parameter server exists), 
doubly-stochastic information mixing matrices are critical in ensuring equal contribution among clients. Generally, obtaining doubly-stochastic matrices can be challenging.  
Push-sum techniques \cite{kempe2003gossip,spiridonoff2020robust} are widely used to address bias that stems from the lack of doubly-stochastic information mixing matrices.
However, 
clients in our problem are only allowed to communicate with the parameter server, 
rendering direct applications of the techniques impossible.
Our setup is orthogonal to them.

\section{Problem Formulation}
A federated learning system consists of %
one %
parameter server and $m$ clients %
that collaboratively  minimize
\begin{align}
\min\limits_{\x\in\reals^d}F\pth{\x} = \frac{1}{m} \sum_{i\in[m]} F_i \pth{\x}, 
\label{eq: global objective}
\end{align}
where
$F_i\pth{\x} = \Expect_{\xi_i\in \calD_i}[\ell_i \pth{\x;\xi_i}]$ is the local objective, 
$\calD_i$ is the local distribution, %
$\xi_i$ is a stochastic sample that client $i$ has access to,
and $\ell_i$ is the local loss function.
The loss function can be non-convex. 

We are interested in solving Eq.\,\eqref{eq: global objective} over unreliable communication uplinks between the parameter server and the clients. 
In each round $t$, the communication uplink between the parameter server and the client $i$ is active with probability $p_i^t$, which could be simultaneously {\em time-varying} and is {\em unknown} to both parameter server and clients.  
We assume that $p_i(t)\ge c$ for all $t$ and all $i$, where $c\in (0,1]$ is an absolute constant. Intuitively, $c$ can be interpreted as one of 
the system configurations. 
For our algorithm to work,
{\em neither} the parameter server {\em nor} clients are required to know $c$.
We use $\calA^t$ to denote the set of clients with active uplinks in round $t$.  

\vskip 0.5\baselineskip
\noindent{\bf Notations.} 
We introduce the additional notations that we will use throughout the paper.
For a given vector $\bm{v}$, $\norm{\bm{v}}$ defines its $l_2$ norm.
For a given matrix $A$, $\fnorm{A}$ defines its Frobenius norm, and 
$\lambda_2(A)$ denotes its second largest eigenvalue when $A$ is a square matrix.  
$\reals^d$ defines a $d$-dimensional vector space. %
$[m] \triangleq \{1, \cdots, m\}$. %
$\indc{\calE}$ is an indicator function of event $\calE$, \ie, $\indc{\calE}=1$ when the event $\calE$ occurs; $\indc{\calE}=0$ otherwise.
$\calF^t$ denotes the sigma-algebra generated by all the randomness up to round $t$. 
$O(\cdot)$ is the asymptotic upper bound of a function growth,  \ie, $f(n) = O(g(n))$ if there exist constants $c_0>0$ and $n_0 \in \naturals$ such that $f(n)\le c_0 g(n)$ for all $n \ge n_0$.

\section{A Case Study on the Bias of {FedAvg}}
\label{sec: case study counterexample}
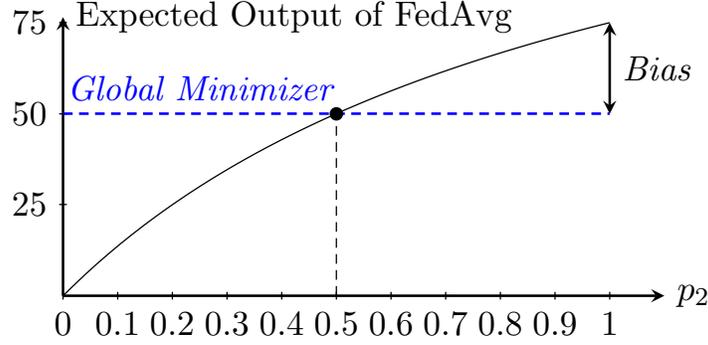
\begin{figure}[!tb]
\centering
\resizebox{.6\columnwidth}{!}{
\begin{tikzpicture}[scale=6] %
    \draw[-stealth,thick] (0,0) -- (1.1,0) node[right] {$p_2$};
    \draw[-stealth,thick] (0,0) -- (0,0.51) node[above, anchor=west] {Expected Output of FedAvg};
    \foreach \x in{0,0.1,0.2,0.3,0.4,0.5,0.6,0.7,0.8,0.9,1} \draw (\x,.2pt) -- (\x,-.2pt) node[anchor=north] {\x};
    \foreach \y/\yt in {0.167/25, 0.333/50, 0.5/75} \draw (.2pt,\y) -- (-.2pt,\y) node[anchor=east] {\yt};
    \draw[densely dashed] (0.5,0) -- (0.5,1/3) node[anchor=south] {};
    \draw[blue,densely dashed, thick] (0,1/3) -- (1,1/3) node[anchor=west,pos=0.25, above] {\it Global Minimizer};
    \draw[black, thick, fill=black] (0.5,0.333) circle (.01cm);
    \draw[stealth-stealth,thick] (1.,0.333) -- (1.,0.5) node[anchor=west,midway] {\it Bias};
    \draw[domain=0:1,smooth,variable=\x] plot ({\x},{(150*\x/(\x+1))/75/2});
\end{tikzpicture}}
\caption{\small
A visualization of the expected output of FedAvg algorithm with two clients, whose $u_1 = 0, u_2 = 100$ and $p_1=0.5$.
We vary $p_2\in [0,1]$ (shown as $x$-axis). %
Eq.~\eqref{eq: bias optimum} becomes %
$\lim_{T\diverge} \expect{x^T} =
\pth{150 \cdot p_2}/\pth{ p_2 + 1}$.
$y$-axis is the expected output of FedAvg.
When $p_2=0.5$, FedAvg recovers the global minimizer $(u_1+u_2)/2 = 50$.
It can be seen that the expected output of the FedAvg algorithm can deviate far from the global minimizer when $p_1 \neq p_2$.
}
\label{fig: number line concrete example}
\end{figure}

In this section, we use a simple quadratic counterexample (a similar setup as in \cite{wang2020tackling}) to illustrate {FedAvg} 
fails to minimize the global objective in 
Eq.\,\eqref{eq: global objective} when $p_i$'s vary across clients.   
Let the local objective $F_i\pth{\x} = \frac{1}{2} \norm{\x - \bu_i}^2,$ where $\bu_i\in \reals^d$ is an arbitrary vector. The corresponding global objective is thus 
\begin{align}
F\pth{\x} = \frac{1}{m} \sum_{i=1}^m F_i\pth{\x} = \frac{1}{2m} \sum_{i=1}^m \norm{\x - \bu_i}^2,\label{eq: counterexample global objective}
\end{align}
with unique 
minimizer %
$\x^\star = \frac{1}{m} \sum_{i=1}^m \bu_i.$ 
\begin{proposition}
\label{proposition: nonuniform}
Choose $\x^0 = \bm{0}$ and $\eta_t = \eta \in (0,1)$ for all $t$. 
For a global objective as per Eq.\,\eqref{eq: counterexample global objective} when $p_i^t=p_i$ for all $t$, 
it holds,
under {FedAvg} with exact local gradients,
that %
\begin{align}
\nonumber
&\lim\limits_{T\diverge}\expect{\x^T} \\
\label{eq: bias optimum}
&~= \sum_{i=1}^m\frac{p_i\bu_i \qth{1
+\sum_{j=2}^{m} \pth{-1}^{j+1} \frac{1}{j}\sum_{ 
S \in \calB_j} \prod_{z\in S} p_z}}{1-\prod_{i=1}^m \pth{1-p_i}},  
\end{align}
where $\calB_j \triangleq \sth{S\Big |S\subseteq [m]\setminus\sth{j}, \abth{S} = j-1}.$
\end{proposition}
It can be checked that if there exist $i, i^{\prime}\in [m]$ such that $p_i\not=p_{i^{\prime}}$,  then $\lim_{t\diverge}\expect{\x^t}\not= \x^*$. 
In fact, the output of FedAvg 
may be significantly away from $\x^{\star}$ depending on $p_i$'s and $\bu_i$'s. 
As illustrated in the scalar example in Fig.\,\ref{fig: number line concrete example},  
overall, 
the global model in FedAvg  %
deviates away from the global optimum.
It is easy to see that the bias only worsens when the connection probabilities $p_i$'s change over time.

\section{Algorithm: Federated Postponed Broadcast (FedPBC)} 
\label{sec: algorithm}
In this section, we propose FedPBC ({\em Federated Postponed Broadcast}, formally described in Algorithm \ref{alg: fedavg variant}) - a simple variant of {FedAvg}. 
Recall that $\calA^t$ denotes all clients with active communication links in global round $t$.  The stochastic gradient used by client $i$ round $t$ is denoted as $\nabla \ell_i(\x_i^{(t,k)}; \xi_i^t)$. 

\begin{algorithm}[!t]
\textbf{Input:} $T$, $\x^0$, $s$, $\sth{\eta_t}_{t=0, \cdots, T-1}$.
The parameter server and each client initialize parameter $\x^0$; 

\For{$t=0, \cdots, T-1$}
{
\tcc{\color{blue} On the clients.}
\For{$i\in [m]$}
{
$\x_i^{(t,0)} = \x_i^{t}$\; 

\For{$k=0, \cdots, s-1$}
{
$\x_i^{(t, k+1)} \gets \x_i^{(t, k)} - \eta_t \nabla \ell_i(\x_i^{(t, k)}; \xi_{i}^t)$\;  
}

$\x_i^{t\star} \gets \x_i^{(t,s)}$\;

Report $\x_i^{t\star}$ to the parameter server\; 
} 
\tcc{\color{blue} On the parameter server.}
\lIf{$\calA^t\not=\emptyset$}{$\x^{t+1} \gets \frac{1}{\abth{\calA^t}} \sum_{i\in \calA^t}\x_i^{t\star}$}
\lElse{
$\x^{t+1} \gets \x^{t}$
}
\lFor{$i\in \calA^t$}
{
$x_i^{t+1} \gets \x^{t+1}$ 
}
}
\caption{FedPBC}
\label{alg: fedavg variant}
\end{algorithm}

Compared to FedAvg,
FedPBC postpones the global model broadcasts to clients in $\calA^t$ till the end of each round. 
Postponing the global model broadcast introduces some staleness as the clients will start from different $\x_i^t$ rather than $\x^t$. 
It turns out that such staleness helps in mitigating the bias caused by non-uniform link activation probabilities.
Moreover, the expected staleness is bounded as shown in~\prettyref{prop: staleness term}.
Theoretical analysis and numerical results can be found in Sections \ref{sec: convergence results} and \ref{sec: numerical}, respectively.

\vskip 0.6\baselineskip 
\noindent{\bf Implicit gossiping among clients in $\calA^t$. }
From line 11 to line 13 of Algorithm \ref{alg: fedavg variant}, via the coordination of the parameter server, the clients in $\calA^t$ {\em implicitly} average their local updates with each other, i.e., there is implicit gossiping among the clients in $\calA^t$ at round $t$. 
Formally, we are able to construct a mixing matrix $W^{(t)}$ as
\begin{align}
\label{eq: gossiping matrix}
W_{ij}^{(t)} =
\begin{cases} 
\frac{1}{\abth{\calA^t}}, &~~~~ \text{if }i, j\in \calA^t; \\    
1, &~~~~ \text{if } i=j\,\text{and} \,\sth{i\notin \calA^t};\\ %
0, &~~~~ \text{otherwise}. 
\end{cases}
\end{align}
The matrix is by definition {\it doubly-stochastic} and $W^{(t)} = \identity$ when $\calA^t =\emptyset$ or $|\calA^t| =1$.   
We further note that this matrix can be {\it time-varying} since the link activation probabilities $p_i^t$'s can be {\it time-varying}. 
As can be seen later, %
this mixing matrix bridges the gap between local and global model heterogeneity and establishes a consensus among clients.
In matrix form, we adopt the following notations. 
\begin{align*}
\bm{X}^{(t)} & = \qth{\x_1^t, \cdots, \x_m^t};   \\
\bm{G}_0^{(t)} &= \qth{s \nabla \ell_1(\x_1^{(t,0)}), \cdots, s \nabla \ell_m(\x_m^{(t,0)})};\\
\bm{G}^{(t)} & = \qth{\sum_{r=0}^{s-1}\nabla \ell_1(\x_1^{(t,r)}), \cdots,  \sum_{r=0}^{s-1}\nabla \ell_m(\x_m^{(t,r)})} ;\\
\nabla \bm{F}^{(t)} & = \qth{\nabla F_1(\x_1^t), \cdots,  \nabla F_m(\x_m^t)}. 
\end{align*}
Consequently,
the consensus error, 
which measures the distance between the averaged model over all the clients and %
local models, 
can be written in matrix form as \eqref{eq: consensus writeup},
\begin{align}
\nonumber
&\frac{1}{m}\sum_{i=1}^m \norm{\bar{\x}^{t} - \x_i^t}^2 
\triangleq \frac{1}{m}
\fnorm{\bm{X}^{(t)} \pth{\identity - \allones}}^2\\
\nonumber
&\qquad \qquad
= \frac{1}{m}
\fnorm{\pth{\bm{X}^{(t-1)} - \eta \bm{G}^{(t-1)}} W^{(t-1)} \pth{\identity - \allones}}^2 \\
\label{eq: consensus writeup}
&\qquad \qquad
= \frac{\eta^2}{m} \fnorm{\sum_{q=0}^{t-1} \bm{G}^{(q)} \pth{\prod_{l=q}^{t-1} W^{(q)} - \allones}}^2,
\end{align}
where the last equality follows from the fact that all clients are initiated at the same weights.

\section{Convergence Analysis}
\label{sec: convergence results}
\subsection{Assumptions}
Before diving into our convergence results, 
we introduce the regularity assumptions, 
which are commented towards the end of this subsection. 
\begin{assumption}[Smoothness]
\label{ass: 2 smmothness}
Each local gradient function
$\nabla \ell_{i}(\theta)$ is $L_i$-Lipschitz, \ie,
$$
\norm{\nabla \ell_{i}(\x_1)-\nabla \ell_{i}(\x_2)}
\le L_i \norm{\x_1-\x_2}
\le L \norm{\x_1-\x_2}
,
$$
for all $\x_1, \x_2,$ and $i\in[m]$,
where $L \triangleq \max\limits_{i\in[m]} L_i$. 
\end{assumption}
\begin{assumption}[Bounded Variance]
\label{ass: bounded variance client-wise}
Stochastic gradients at each client node $i\in[m]$ are unbiased estimates of the true gradient of the local objectives, i.e.,  
\[
\expect{\nabla \ell_i(\x_i^t) \mid \calF^{t}}=\nabla F_i(\x_i^t),  
\]
and the variance of stochastic gradients at each client node $i\in[m]$ is uniformly bounded, i.e., 
\[
\expect{\norm{\nabla \ell_i(\x)-\nabla F_i(\x)}^2 \mid \calF^t}\le\sigma^2.
\]
\end{assumption}
\begin{assumption}
\label{ass: lower bounds}
There exists $F^*\in \reals$ such that $F(\x)\ge F^*$ for all $\x\in \reals^d$. 
\end{assumption}
\begin{assumption}[Bounded Inter-client Heterogeneity]
\label{ass: bounded similarity}
We say that local objective function $F_i$'s satisfy $(\beta,\zeta)$-bounded dissimilarity condition for $\beta,\zeta \ge 0$ if
\begin{align}
\label{eq: BG condition}
\frac{1}{m}\sum_{i=1}^m \norm{\nabla F_i(\x)- \nabla F(\x)}^2 \le \beta^2 \norm{\nabla F(\x)}^2+ \zeta^2. 
\end{align}
\end{assumption}

Assumptions, \ref{ass: 2 smmothness},  \ref{ass: bounded variance client-wise} and \ref{ass: lower bounds} are standard in federated learning analysis \cite{karimireddy2020scaffold,li2020federated,yuan2022}.
Assumption \ref{ass: bounded similarity} captures the heterogeneity across different users.
It is a more relaxed assumption, \eg than, bounded gradients \cite{cho2022towards,yan2023federated}, where they assume $\frac{1}{m}\sum_{i \in [m]} \norm{\nabla F_i(\x)}^2 \le \zeta^2$, also than \cite{wang2022,yang2022anarchic}, where they assume $\frac{1}{m}\sum_{i \in [m]} \norm{\nabla F_i(\x) - \nabla F(\x)}^2 \le \zeta^2$.
When clients have \iid local datasets, \ie, $\calD_i = \calD_j$ for $\forall~i,j\in[m]$,
it holds for Eq.~\eqref{eq: BG condition} that $\beta = \zeta=0$ since $F_i = F_j$.
Notably, 
we assume the unbiasedness in Assumption \ref{ass: bounded variance client-wise} is imposed only at the beginning of each global round. 

\subsection{Convergence Results}
In this section,
we state our key lemmas and our main theorem. 
The proofs of Lemma~\ref{lemma: descent lemma} and \ref{lemma: consensus} are deferred to~\prettyref{sec: selected proofs}.
All remaining proofs are relegated to~\prettyref{app: proofs}.
\prettyref{prop: staleness term} 
captures 
the expected staleness of local updates. %
\begin{proposition}
\label{prop: staleness term}
Define the last active round of the link $i$ as $\tau_i(t) \triangleq \{t^\prime \mid t^\prime < t, i \in \calA^{t^\prime}\}$.
Given $p_i^t$ such that $p_i^t \ge c$, where $c$ is an absolute constant, %
we have $\expect{ t - \tau_i(t)}\le \frac{1}{c}$. 
\end{proposition}
\begin{lemma}[Lemma 1 in \cite{su2023federatedadv}]
\label{lemma: local step perturbation}
For $s\ge 1$, 
suppose~\prettyref{ass: 2 smmothness} holds,
we have for all $\x\in\reals^d:$

\[
\norm{\sum_{k=0}^{s-1}\qth{ \nabla \ell_i (\x^{\pth{t,k}}) - \nabla \ell_i (\x^t)}}
\le \kappa \eta \binom{s}{2} L_i \norm{\nabla \ell_{i}(\x^t)},
\]

where
$
\kappa \triangleq \max_{i}\frac{(1+\eta L_i)^s - 1- s \eta L_i}{\binom{s}{2}\pth{\eta L_i}^2} 
$
and monotonically non-decreases with respect to $\eta>0$.
\end{lemma}
\begin{remark}
\label{rmk: step perturbation}
Lemma~\ref{lemma: local step perturbation} comes from a concurrent work \cite{su2023federatedadv} and characterizes the perturbation incurred by the multi-step local computation.
When $s=1$, \ie, 
when a client performs only one-step local computation,
it holds that $\kappa = 0$.
For $s \ge 2$, 
we have $\kappa \ge 1$.
Moreover, 
due to its monotonicity with respect to $\eta$ in Lemma~\ref{lemma: local step perturbation}, 
$\kappa$ is bounded from above by an absolute constant when the learning rate $\eta$ is upper bounded. 
Let 
\end{remark}
\begin{align}
\label{eq: average estimate}
\bar{\x}^t \triangleq \frac{1}{m}\sum_{i=1}^m \x_i^t. 
\end{align}
\begin{lemma}[Descent Lemma]
\label{lemma: descent lemma}
Suppose Assumptions \ref{ass: 2 smmothness}, \ref{ass: bounded variance client-wise}, and \ref{ass: bounded similarity} hold.
Choose a learning rate $\eta$ such that
$\eta \le \frac{1}{108L^2 s^3 (\beta^2+1)(1+\kappa^2L^2)}$.
When Lipschitz constant $L \ge 1$,
it holds that
\begin{align*}
&\expect{F(\bar{\x}^{t+1})  -  F(\bar{\x}^{t}) \mid \calF^{t}}
\le -\frac{\eta s}{3} \norm{\nabla F(\bar{\x}^t)}^2 \\
&~~~+ \eta s\frac{L^2}{m} \sum_{i=1}^m \norm{\x_i^t - \bar{\x}^t}^2 
+ \eta^2 s^2 6L\pth{\zeta^2+\sigma^2}\pth{1+\kappa^2 L^2}. 
\end{align*} 
\end{lemma}

The consensus error term $\frac{1}{m} \sum_{i=1}^m \norm{\x_i^t - \bar{\x}^t}^2$ %
in Lemma~\ref{lemma: descent lemma}
connects our analysis to the aforementioned $W$ matrix.
Let
\begin{align*}
&M^{(t)}\triangleq \expect{\pth{W^{(t)}}^2}, \quad
\allones \triangleq \frac{1}{m} \Indc \Indc^\top; \\ 
&\rho(t)\triangleq\lambda_2\pth{M^{(t)}}
~~~\text{and}~
\rho \triangleq \max_t \rho(t).
\end{align*}

Next,
we borrow insights from the analysis of gossiping algorithms in the following lemma.

\begin{lemma}[Ergodicity]
\label{lemma: ergodicity}
If $p_i^t\ge c$ for some constant $c\in (0,1)$. 
\begin{itemize}
\item 
For each $t\ge 1$, it holds that 
$\rho \le 1 - \frac{c^4\qth{1-\pth{1-c}^m}^2}{8}$;
\item 
In the special case of uniform and time-invariant availability,
suppose it holds that $\abth{\calA^t} = k$ for all $t \ge 0$,
the bound can be further tightened as
$\rho \le 1 - \frac{c^2}{8}$,
where $c \triangleq k/m$.
\end{itemize}

\textnormal{(Mixing rate, \cite[Lemma 1]{wang2022matcha})}{\bf .}
For any matrix $B\in \reals^{d\times m}$, 
it holds that
\begin{align}
\label{eq: mathcha consensus}
\expects{\fnorm{B\pth{\prod_{r=1}^t W^{(r)} - \allones}}^2}{W} 
\le 
\rho^t\fnorm{B}^2,
\end{align}
where $\expects{\cdot}{W}$ denotes an expectation taken with respect to randomness in $W^{(1)}, \cdots, W^{(t)}$.
\end{lemma}
Inequality~\eqref{eq: mathcha consensus} from \cite[Lemma 1]{wang2022matcha} 
enables us to bound the consensus error term $\frac{1}{m} \sum_{i=1}^m \norm{\x_i^t - \bar{\x}^t}^2$ and 
says that the spectral norm $\rho$ must be less than 1 to ensure a bounded error, which is crucial for the objectives to reach a stationary point.
Fortunately,
we show that,
under our uplink availability assumption,
$\rho < 1$ in Lemma~\ref{lemma: ergodicity}.
\begin{lemma}[Consensus Error]
\label{lemma: consensus}
Suppose Assumptions \ref{ass: 2 smmothness}, 
\ref{ass: bounded variance client-wise}, and \ref{ass: bounded similarity} hold.
Choose a learning rate $\eta$ such that
$\eta \le \frac{1-\sqrt{\rho}}{108 L^2 s^3 \pth{\beta^2+1} \pth{1+\kappa^2 L^2}}$.
When Lipschitz constant $L \ge 1$,
it holds that
\begin{align*}
&\frac{1}{mT}\sum_{t=0}^{T-1}\expect{\fnorm{\bm{X}^{\pth{t}} \pth{\identity - \allones}}^2} 
\le 
\frac{12\rho \sigma^2}{(1-\sqrt{\rho})^2} \eta^2 s^2 \\
&~+\frac{54\rho \zeta^2}{(1-\sqrt{\rho})^2} \eta^2 s^2
+\frac{54(\beta^2+1)\rho \eta^2 s^2}{(1-\sqrt{\rho})^2} \frac{1}{mT} \sum_{t=0}^{T-1} \fnorm{\nabla F(\bar{\bm{x}}^t)}^2.
\end{align*}
\end{lemma}
Our proof of Lemma~\ref{lemma: consensus} shares a similar sketch as that in \cite{wang2022matcha} yet with non-trivial adaptation to account for multiple local updates and the fact the stochastic gradients at a client within each round are {\em not independent}. 
Plugging Lemma~\ref{lemma: consensus} into Lemma~\ref{lemma: descent lemma},
we obtain the main~\prettyref{thm: main}.
\begin{theorem}
\label{thm: main}
Suppose Assumptions \ref{ass: 2 smmothness}, %
\ref{ass: bounded variance client-wise}, 
\ref{ass: lower bounds}, %
and \ref{ass: bounded similarity} %
hold. 
Choose a learning rate $\eta $ such that %
$\eta \le \frac{1-\sqrt{\rho}}{108 L^2 s^3 \pth{\beta^2+1} \pth{1+\kappa^2 L^2}}$. 
When Lipschitz constant $L \ge 1$,
it holds that
\begin{align*}
\frac{1}{T}\sum_{t=0}^{T-1}
&\expect{\norm{\nabla F(\bar{\x}^t)}^2}
\le
\frac{6 \pth{F(\bar{\x}^{0})- F^\star }}{ \eta s T} \\
&~~~+
54 \eta s L
\pth{
\kappa^2 L^2 + 1
+
\frac{1}{1 - \sqrt{\rho}}}
\pth{\sigma^2 + \zeta^2}
.
\end{align*}
\end{theorem}
\begin{corollary}
\label{crl: main}
Suppose Assumptions \prettyref{ass: 2 smmothness}, %
\ref{ass: bounded variance client-wise}, %
\ref{ass: lower bounds}, %
and \ref{ass: bounded similarity} %
hold. 
Choose %
$\eta = 1/\sqrt{T}$ %
such that
$\eta \le \frac{1-\sqrt{\rho}}{108 L^2 s^3 \pth{\beta^2+1} \pth{1+\kappa^2 L^2}}$. 
When Lipschitz constant $L \ge 1$,
it holds that
\begin{align*}
\frac{1}{T}\sum_{t=0}^{T-1}
&\expect{\norm{\nabla F(\bar{\x}^t)}^2}
\le
\frac{6 \pth{F(\bar{\x}^{0})- F^\star }}{ s \sqrt{T}} \\
&~~~+
54 \frac{s L}{\sqrt{T}}
\pth{
\kappa^2 L^2 + 1
+
\frac{1}{1 - \sqrt{\rho}}}
\pth{\sigma^2 + \zeta^2}
.
\end{align*}
\end{corollary}
\begin{remark}
Here, we remark on Theorem \ref{thm: main}: \\
(1) {\bf On the structures.} 
The assumption that Lipschitz constant $L \ge 1$ is for simplifying the upper bound of $\eta$ only,
which, notably,
can be readily relaxed but at a cost of a much more sophisticated learning rate condition.
The second term stems from noisy stochastic gradients (\prettyref{ass: bounded variance client-wise}) and inter-client gradient heterogeneity (\prettyref{ass: bounded similarity}).\\  %
(2) {\bf On stationary points of $F$.} 
\prettyref{thm: main} says that $\bar{\x}^t$ in FedPBC converges to a stationary point of $F$ (non-convex) at a rate of $1/\sqrt{T}$. 
In sharp contrast, 
Proposition \ref{proposition: nonuniform} dictates that the expected output of FedAvg converges to a point that could be far away from the true optimum depending on the interplay between $p_i^t$'s and data heterogeneity. \\
(3) {\bf On the role of the probability lower bound $c$.} 
 A larger $c$ results in a smaller $\rho$ and thus a tighter bound on $\frac{1}{T}\sum_{t=0}^{T-1} \expect{\norm{\nabla F \pth{\bar{\x}^t}}}.$ 
Next,
we discuss a couple of special cases 
in Big-O notation with respect to
the number of clients $m$,
the number of local steps $s$,
spectral norm $\rho$,
stochastic gradient variance $\sigma$
and bounded gradient dissimilarity $\zeta$.
\begin{itemize}[leftmargin=*]
\item
FedPBC reduces to FedAvg with full-client participation when $c=1$.  %
Setting $\eta = \sqrt{m/sT}$ in~\prettyref{thm: main},
our convergence rate $O(\frac{1}{\sqrt{msT}} + \sqrt{\frac{ms}{T}}\pth{\sigma^2 + \zeta^2})$ matches the FedAvg literature  
(\eg, \cite{wang2020tackling}). 
\item
When it comes to FedAvg with uniform and time-invariant participation,
suppose $k$ out of $m$ clients are selected uniformly at random each round.
Setting $\eta = \sqrt{k/sT}$ in~\prettyref{thm: main},
our convergence rate becomes $O(\frac{1}{\sqrt{ksT}} + \frac{1}{1 - \sqrt{\rho}} \sqrt{\frac{ks}{T}}\pth{\sigma^2 + \zeta^2})$.
Since
$\rho \le 1 - {c^2}/{8}$ %
(in Lemma~\ref{lemma: ergodicity}),
the rate becomes 
$O(\frac{1}{\sqrt{ksT}} + \frac{1}{c^2} \sqrt{\frac{ks}{T}}\pth{\sigma^2 + \zeta^2}),$    
which %
introduces a larger variance compared to the rate of FedAvg with full participation,
consistent with existing literature
(\eg, \cite{yang2021achieving}).
\end{itemize} 
(4) {\bf On convergence rate.} 
Our convergence rate in Corollary~\ref{crl: main} of $O(1/\sqrt{T})$,
where the Big-O notation is taken with respect to the total global round $T$,
matches the best possible rate for any first-order algorithms that have access to only noisy stochastic gradients of a smooth non-convex objective \cite{nemirovski2009robust}.
By setting learning rate $\eta$ as in bulletpoint (3),
we shall see linear speedup with respect to the first term; however, the second term ultimately dominates the first term, which is consistent with {FedAvg} literature, 
see, \eg, \cite{Li2020}. 
We leave a future direction to achieve linear speedup. 
\end{remark}

\section{Numerical Experiments}
\label{sec: numerical}
In this section, 
we evaluate FedPBC and multiple baseline algorithms on a simple quadratic function and real-world datasets.
\subsection{Quadratic function}
\begin{figure}[!t]
\centering
\vspace{-1.5em}
\includegraphics[width=\columnwidth]{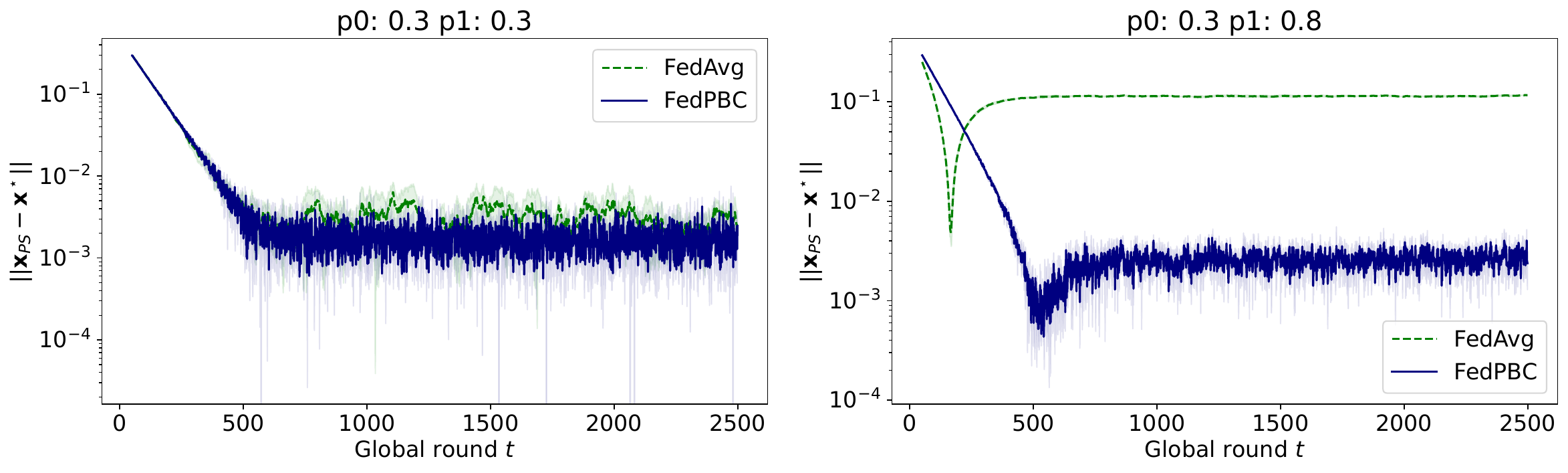}
\vskip -1pt
\caption{\small 
$\norm{\x_{\text{PS}} - \x^\star}$ %
in logarithmic scale.
The results are obtained after an average of $3$ random seeds.
Plots are reported as mean $\pm$ standard deviation.
The shaded areas plot standard deviation.
}
\label{fig: counterexample}
\end{figure}

The first part is about a simple quadratic function as in Eq.~\eqref{eq: counterexample global objective}.
Recall that,
in each round $t$, 
client $i$ responds to the parameter server's update request with probability $p_i^t$.
\newline
\noindent{\bf Counterexample.} 
Our numerical results can be found in~\prettyref{fig: counterexample}. 
We consider a federated learning system of $m = 100$ clients, 
each performing $s=100$ steps local updates per round, %
in a total of $2500$ global rounds.
The local objective %
is 
$F_i(\x_i) = \frac{1}{2}\norm{\x_i - \bu_i}^2,$
where $\x_i, \bu_i \in \reals^{100},$ 
$\bu_i\sim \calN \pth{ (i/1000) \Indc, 0.01 \identity },$
and $\x_i^0=\bm{0}$
for all $i\in[m].$
The learning rate $\eta = 0.0001$. %
The uplinks of the first $50$ clients become open with probability $p_0,$ whereas the second half with $p_1$  -- to be specified later. 
For ease of presentation, we plot the distance to the optimum $\norm{\x_{\text{PS}} - \x^\star}$ after the first $50$ communication rounds in Fig.\,\ref{fig: counterexample}, where $\x_{\text{PS}}^t \triangleq \x^t$ in Algorithm \ref{alg: fedavg variant}.
All results are obtained after $3$ random seeds and reported as mean $\pm$ standard deviation.
Notably,
all plots are on a logarithmic scale, 
potentially magnifying visual fluctuations.
Notice that the distance to optimum $\norm{\x_{\text{PS}} - \x^\star}$ does not go strictly to 0.
We presumably attribute this to pseudo-randomness in computers to sample clients.
Observe that two algorithms attain a similar distance to optimum when $p_0 = p_1$. Yet,
FedPBC obtains a much lower error when $p_0 \neq p_1$.
In addition,
the error is on a similar scale (around $10^{-3}$) as in the case of $p_0 = p_1$.

\subsection{Real-world Datasets}
\label{sec: real world numerical}
\FloatBarrier
\begin{table}[H]
\centering
\caption{\small 
The reported results are in the form of mean accuracy $\pm$ standard deviation and are obtained over 3 repetitions in different random seeds.
Results are averaged over the last $100$ rounds.
In each simulation, clients perform mini-batch stochastic gradient descent in $5$ steps on a convolutional neural network (CNN) locally per round.
The total global rounds for SVHN, CIFAR-10, CINIC-10 are 4000, 10000, 10000, respectively.
Furthermore, we use customized CNNs for different datasets, respectively.
Algorithms are categorized into two groups:
(1) ones {\em not} aided by memory or known statistics;
(2) ones with memory (including MIFA and FedAvg with {\it known} $p_i^t$'s).
Moreover, we highlight the best and the second best in \colorbox{yellow!50}{yellow} and in \colorbox{cyan!10}{cyan}, respectively, among algorithms {\em not} aided by memory or known statistics.
The other hyperparameters are specified in~\prettyref{app: hyperparameter}, 
and some of them are tuned using grid search.
}
\label{tbl: exp main text}
\resizebox{\linewidth}{!}{
\begin{tabular}{c|c|cc|cc|cc}
    \toprule
    \multirow{2}{*}{\begin{tabular}{@{}c@{}}{\bf Unreliable} \\ {\bf Patterns} \end{tabular}} &
    {\bf Datasets} &
    \multicolumn{2}{c|}{SVHN} & 
    \multicolumn{2}{c|}{CIFAR-10} & 
    \multicolumn{2}{c}{CINIC-10} \\
    \cline{2-8}
    & 
    {\bf Algorithms}&
    Train &
    Test &
    Train &
    Test &
    Train &
    Test  \\
    \hline
     \multirow{6}{*}{
\begin{tabular}{@{}c@{}} Bernoulli$^{1}$ \\ with {\em time-invariant} $p_i$'s \\ 
\end{tabular}} & 
FedPBC ({ours})
& 
\cellcolor{yellow!50}
${84.4}\% \pm 0.008$ &
\cellcolor{yellow!50}
{${84.3}\% \pm 0.008$} &
\cellcolor{yellow!50}
{${68.4}\% \pm 0.011$} &
\cellcolor{yellow!50}
${66.3}\% \pm 0.013$ &
\cellcolor{yellow!50}
{${50.3}\% \pm 0.005$} &
\cellcolor{yellow!50}
{${49.7}\% \pm 0.004$} 
\\
& 
FedAvg & 
$75.9\% \pm 0.024 $ &
$75.2\% \pm 0.024$ &
$59.9\% \pm 0.026 $ &
$58.7\% \pm 0.025$ &
$38.1\% \pm 0.031$ &
$37.8\% \pm 0.029$ 
\\
& 
FedAvg {\em all} & 
$56.4\% \pm 0.083$ &
$56.4\% \pm 0.072$ &
$48.9\% \pm 0.031$ &
$48.7\% \pm 0.026$ &
$32.6\% \pm 0.030$ &
$32.3\% \pm 0.030$ 
\\
& 
FedAU & 
\cellcolor{cyan!10}
${83.1}\% \pm 0.015$ &
\cellcolor{cyan!10}
$83.0\% \pm 0.015$ &
\cellcolor{cyan!10}
$67.4\% \pm 0.019 $ &
\cellcolor{cyan!10}
$65.9\% \pm 0.019$ &
\cellcolor{cyan!10}
$45.8\% \pm 0.022$ &
\cellcolor{cyan!10}
$45.4\% \pm 0.022$ 
\\
& 
F3AST & %
$76.9\% \pm 0.036$ &
$76.9\% \pm 0.037$ &
$58.5\% \pm 0.053$ &
$57.7\% \pm 0.052$ &
$40.7\% \pm 0.049$ &
$40.3\% \pm 0.048$ 
\\

\noalign{\vspace{.5mm}}
\cline{2-8}
\noalign{\vspace{.5mm}}
& 
FedAvg {\em known} $p_i$'s & 
$77.8\% \pm 0.029$ &
$77.2\% \pm 0.032$ &
$61.1\% \pm 0.036$ &
$60.1\% \pm 0.035$ &
$39.2\% \pm 0.029$ &
$38.8\% \pm 0.029$ 
\\
& 
MIFA ({\em memory aided}) & 
${80.8}\% \pm 0.003$ &
${80.8}\% \pm 0.003$ &
${67.8}\% \pm 0.006 $&
${67.1}\% \pm 0.006$&
${47.6}\% \pm 0.005$ &
${47.1}\% \pm 0.005$ \\
    \hline
    \multirow{6}{*}{
\begin{tabular}{@{}c@{}} Bernoulli \\ with {\em time-varying} $p_i^t$'s \end{tabular}} & 
FedPBC ({ours})& 
\cellcolor{yellow!50}
{${84.0\%} \pm 0.009$} &
\cellcolor{yellow!50}
{${84.0\%} \pm 0.009$} &
\cellcolor{yellow!50}
{${67.1\%} \pm 0.011$} &
\cellcolor{yellow!50}
${65.0}\%\pm 0.015$ &
\cellcolor{yellow!50}
{${49.7\%} \pm 0.004$} &
\cellcolor{yellow!50}
{${49.1\%} \pm 0.003$} 
\\
& 
FedAvg & 
$73.7\% \pm 0.041 $ &
$72.7\% \pm 0.042$ &
$57.3\% \pm 0.034 $ &
$56.2\% \pm 0.033$ &
$35.9\% \pm 0.038$ &
$35.6\% \pm 0.037$ 
\\
& 
FedAvg {\em all} & 
$37.0\% \pm 0.097$ &
$36.5\% \pm 0.085$ &
$43.2\% \pm 0.030$ &
$43.2\% \pm 0.029$ &
$28.9\% \pm 0.024$ &
$28.7\% \pm 0.024$ 
\\
& 
FedAU & 
\cellcolor{cyan!10}
${80.5}\%\pm 0.023$ &
\cellcolor{cyan!10}
${80.3}\%\pm 0.022$ &
\cellcolor{cyan!10}
$64.9\% \pm 0.018 $ &
\cellcolor{cyan!10}
$63.5\% \pm 0.018$ &
\cellcolor{cyan!10}
$44.8\% \pm 0.017$ &
\cellcolor{cyan!10}
$43.4\% \pm 0.018$ 
\\
& 
F3AST & %
$78.3\% \pm 0.027$ &
$78.1\% \pm 0.029$ &
$60.7\% \pm 0.037$ &
$59.6\% \pm 0.035$ &
$41.2\% \pm 0.035$ &
$40.8\% \pm 0.035$ 
\\

\noalign{\vspace{.5mm}}
\cline{2-8}
\noalign{\vspace{.5mm}}
& 
FedAvg {\em known} $p_i^t$'s 
& 
$76.9\% \pm 0.035$ &
$76.3\% \pm 0.036$ &
$62.4\% \pm 0.021$ &
$61.2\% \pm 0.022$ &
${46.9}\%\pm 0.016$ &
${46.4}\%\pm 0.016$ 
\\
&
MIFA ({\em memory aided}) & 
${79.2}\% \pm 0.005$ &
${79.2}\% \pm 0.005$ &
${66.2}\%\pm 0.006 $&
${65.5}\% \pm 0.005$&
${46.4}\% \pm 0.010$ &
${45.8}\% \pm 0.009$ \\
    \toprule 
    \multirow{6}{*}{
\begin{tabular}{@{}c@{}}
Homogeneous$^{1}$ \\ Markovian \\ with {\em time-invariant} $p_i$'s \end{tabular}} & 
FedPBC ({ours})& 
\cellcolor{yellow!50}
{${84.8}\% \pm 0.009$} &
\cellcolor{yellow!50}
{${84.1}\% \pm 0.008$} &
\cellcolor{yellow!50}
{${68.6}\% \pm 0.010$} &
\cellcolor{yellow!50}
{${66.5}\% \pm 0.010$} &
\cellcolor{yellow!50}
{${50.0}\% \pm 0.006$} &
\cellcolor{yellow!50}
{${49.5}\% \pm 0.006$} 
\\
& 
FedAvg & 
$74.7\% \pm 0.023$ &
$74.0\% \pm 0.023$ &
$59.1\% \pm 0.022$ &
$57.9\% \pm 0.020$ &
$37.4\% \pm 0.029$ &
$37.1\% \pm 0.029$ 
\\
& 
FedAvg {\em all} & 
$55.1\% \pm 0.073$ &
$55.1\% \pm 0.063$ &
$48.3\% \pm 0.039$ &
$48.0\% \pm 0.034$ &
$31.6\% \pm 0.032$ &
$31.4\% \pm 0.031$ 
\\
& 
FedAU & 
\cellcolor{cyan!10}
${82.7}\% \pm 0.015$ &
\cellcolor{cyan!10}
${82.6}\% \pm 0.013$ &
\cellcolor{cyan!10}
${68.3}\% \pm 0.019$ &
\cellcolor{cyan!10}
${66.4}\% \pm 0.018$ &
\cellcolor{cyan!10}
${47.2}\% \pm 0.019$ &
\cellcolor{cyan!10}
${46.7}\% \pm 0.018$ 
\\
& 
F3AST& %
$75.5\% \pm 0.043$ &
$75.5\% \pm 0.048$ &
$60.3\% \pm 0.035$ &
$59.3\% \pm 0.034$ &
$43.0\% \pm 0.028 $ &
$42.5\% \pm 0.027$ 
\\

\noalign{\vspace{.5mm}}
\cline{2-8}
\noalign{\vspace{.5mm}}
& 
FedAvg {\em known} $p_i$'s & 
$76.0\% \pm 0.025$ &
$75.7\% \pm 0.027$ &
$61.0\% \pm 0.036$ &
$60.0\% \pm 0.034$ &
$40.8\% \pm 0.022$ &
$40.4\% \pm 0.022$ 
\\
& 
MIFA ({\em memory aided}) & 
${81.7}\% \pm 0.006$ &
${81.1}\% \pm 0.004$ &
${66.8}\% \pm 0.006$ &
${65.9}\% \pm 0.006$ &
${46.9}\% \pm 0.007$ &
${46.4}\% \pm 0.007$ 
 \\
    \midrule 
   \multirow{6}{*}{
\begin{tabular}{@{}c@{}}
Non-homogeneous \\ Markovian \\ with {\em time-varying} $p_i^t$'s \end{tabular}} & 
FedPBC ({ours})& 
\cellcolor{yellow!50}
{${83.9\%} \pm 0.010$} &
\cellcolor{yellow!50}
{${83.8\%} \pm 0.008$} &
\cellcolor{yellow!50}
{${67.2\%} \pm 0.009$} &
\cellcolor{cyan!10}
{${64.9\% \pm 0.006}$} &
\cellcolor{yellow!50}
{${49.7\%} \pm 0.004$} &
\cellcolor{yellow!50}
{${49.1\%} \pm 0.004$} 
\\
& 
FedAvg & 
$72.7\% \pm 0.034$ &
$72.2\% \pm 0.035$ &
$59.0\% \pm 0.027$ &
$58.0\% \pm 0.027$ &
$36.7\% \pm 0.031$ &
$36.3\% \pm 0.030$ 
\\
& 
FedAvg {\em all} & 
$38.6\% \pm 0.091$ &
$38.3\% \pm 0.079$ &
$43.7\% \pm 0.026$ &
$43.8\% \pm 0.024$ &
$29.4\% \pm 0.025$ &
$29.2\% \pm 0.024$ 
\\
& 
FedAU & 
\cellcolor{cyan!10}
${80.2}\% \pm 0.020$ &
\cellcolor{cyan!10}
${80.2}\% \pm 0.020$ &
\cellcolor{cyan!10}
${66.4}\% \pm 0.018$ &
\cellcolor{yellow!50}
${65.1}\% \pm 0.018$ &
\cellcolor{cyan!10}
$45.3\% \pm 0.022$ &
\cellcolor{cyan!10}
$44.8\% \pm 0.021$ 
\\
& 
F3AST & %
$77.0\% \pm 0.033$ &
$77.0\% \pm 0.033$ &
$62.8\% \pm 0.032$ &
$61.5\% \pm 0.032$ &
$43.0\% \pm 0.029$ &
$42.6\% \pm 0.028$ 
\\
\noalign{\vspace{.5mm}}
\cline{2-8}
\noalign{\vspace{.5mm}}
& 
FedAvg {\em known} $p_i^t$'s & 
$76.3\% \pm 0.045$ &
$76.3\% \pm 0.045$ &
$60.0\% \pm 0.040$ &
$59.0\% \pm 0.038$ &
$45.1\% \pm 0.032$ &
$44.5\% \pm 0.031$ 
\\
& 
MIFA ({\em memory aided}) & 
${79.2}\% \pm 0.005$ &
${79.1}\% \pm 0.004$ &
${66.3}\% \pm 0.007 $ &
${65.6\%} \pm 0.007$ &
${46.5}\% \pm 0.008 $ &
${46.1}\% \pm 0.008 $  \\
    \toprule
    \multirow{6}{*}{
\begin{tabular}{@{}c@{}} Cyclic$^{1}$ \\ {\em without} periodic reset 
\end{tabular}}& 
FedPBC ({ours}) & 
\cellcolor{yellow!50}
{${84.2}\% \pm 0.010$} &
\cellcolor{yellow!50}
{${84.2}\% \pm 0.009$} &
\cellcolor{yellow!50}
{${67.5}\% \pm 0.015$} &
\cellcolor{yellow!50}
{${65.2}\% \pm 0.017$} &
\cellcolor{yellow!50}
{${49.7}\% \pm 0.008$} &
\cellcolor{yellow!50}
{${49.0}\% \pm 0.007$} 
\\
& 
FedAvg & 
$72.3\% \pm 0.029 $ &
$71.7\% \pm 0.032$ &
$57.0\% \pm 0.028$ &
$56.0\% \pm 0.026$ &
$37.0\% \pm 0.029$ &
$36.6\% \pm 0.029$ 
\\
& 
FedAvg {\em all} & 
$56.4\% \pm 0.078 $ &
$56.4\% \pm 0.070$ &
$48.5\% \pm 0.026$ &
$48.1\% \pm 0.024$ &
$32.2\% \pm 0.028$ &
$31.9\% \pm 0.027$ 
\\
& 
FedAU & 
\cellcolor{cyan!10}
${80.2}\% \pm 0.027$ &
\cellcolor{cyan!10}
${79.8}\% \pm 0.027$ &
\cellcolor{cyan!10}
${64.5}\% \pm 0.024$ &
\cellcolor{cyan!10}
${63.1}\%\pm 0.022$ &
\cellcolor{cyan!10}
${43.3}\% \pm 0.033$ &
\cellcolor{cyan!10}
${42.8}\% \pm 0.032$ 
\\
& 
F3AST& %
$71.5\% \pm 0.042$ &
$71.7\% \pm 0.044$ &
$58.3\% \pm 0.026$ &
$57.3\% \pm 0.028$ &
$40.0\% \pm 0.028$ &
$39.7\% \pm 0.028$ 
\\

\noalign{\vspace{.5mm}}
\cline{2-8}
\noalign{\vspace{.5mm}}
& 
FedAvg {\em known} $p_i$'s$^{2}$ & 
${74.1}\% \pm 0.037 $ &
${73.6}\% \pm 0.038$ &
$58.9\% \pm 0.036$ &
$58.0\% \pm 0.034$ &
$38.1\% \pm 0.042$ &
$37.7\% \pm 0.041$ 
\\
& 
MIFA ({\em memory aided}) & 
{$70.9\% \pm 0.033$} &
{$70.9\% \pm 0.033$} &
${59.1}\% \pm 0.021$ &
${58.7}\% \pm 0.022$ &
${42.3}\% \pm 0.039$&
${41.8}\% \pm 0.038$ \\
    \midrule
    \multirow{6}{*}{
\begin{tabular}{@{}c@{}} Cyclic \\ {\em with} periodic reset \end{tabular}}& 
FedPBC ({ours})& 
\cellcolor{yellow!50}
{${83.8\%} \pm 0.008$} &
\cellcolor{yellow!50}
{${83.7\%} \pm 0.007$} &
\cellcolor{yellow!50}
{${66.3\%} \pm 0.010$} &
\cellcolor{yellow!50}
${64.0}\% \pm 0.012$ &
\cellcolor{yellow!50}
{${49.6\%} \pm 0.004$} &
\cellcolor{yellow!50}
{${49.1\%} \pm 0.004$} 
\\
& 
FedAvg & 
$69.6\% \pm 0.054$ &
$69.0\% \pm 0.058$ &
$56.0\% \pm 0.032$ &
$55.1\% \pm 0.033$ &
$35.4\% \pm 0.027$ &
$35.1\% \pm 0.026$ 
\\
& 
FedAvg {\em all} & 
$34.2\% \pm 0.074$ &
$33.6\% \pm 0.065$ &
$42.5\% \pm 0.026$ &
$42.4\% \pm 0.026$ &
$28.7\% \pm 0.023$ &
$28.5\% \pm 0.023$ 
\\

& 
FedAU & 
\cellcolor{cyan!10}
$77.1\% \pm 0.029$ &
\cellcolor{cyan!10}
$77.1\% \pm 0.029$ &
\cellcolor{cyan!10}
$62.9\% \pm 0.022$ &
\cellcolor{cyan!10}
$61.7\% \pm 0.021$ &
\cellcolor{cyan!10}
$42.6\% \pm 0.020 $ &
\cellcolor{cyan!10}
$42.1\% \pm 0.020$ 
\\
& 
F3AST & %
$75.4\% \pm 0.035$ &
$75.3\% \pm 0.037$ &
$62.3\% \pm 0.041$ &
$61.0\% \pm 0.040$ &
$42.7\% \pm 0.041$ &
$42.2\% \pm 0.040$ 
\\

\noalign{\vspace{.5mm}}
\cline{2-8}
\noalign{\vspace{.5mm}}
& 
FedAvg {\em known} $p_i$'s$^{2}$  & 
$72.7\% \pm 0.049 $ &
$72.1\% \pm 0.052$ &
$60.0\% \pm 0.032$ &
$59.1\% \pm 0.030$ &
$45.5\% \pm 0.029$ &
$45.0\% \pm 0.028$ 
\\

& 
MIFA ({\em memory aided}) & 
${77.6}\% \pm 0.014$ &
${77.3}\% \pm 0.014$ &
${64.8}\% \pm 0.006$ &
${64.3\%} \pm 0.006$ &
${45.6}\% \pm 0.010$&
${45.2}\% \pm 0.010$  \\
    \bottomrule
\end{tabular}}
\vskip .5\baselineskip
\begin{tabular}{p{.95\textwidth}}
{\footnotesize $^{1}$ Bernoulli, homogeneous Markovian with {\em time-invariant} $p_i$'s and cyclic {\em without} periodic restart have been evaluated in \cite{wang2023lightweight} but with different number of clients and $p_i$'s;} \\
{\footnotesize $^{2}$ The known statistics in cyclic patterns are determined by the ratio of active rounds to the whole cyclic period.} 
\end{tabular}
\end{table}
In this section, we use three real-world datasets to validate 
the performance of FedPBC on different uplink unreliable patterns, and to compare with multiple baseline algorithms.
Detailed hardware and software specifications
can be found in~\prettyref{app: exp}.

\noindent{\bf Dataset and data heterogeneity.}
The image classification task is commonly adopted in evaluating the empirical performance of a federated learning system \cite{li2020federated,mcmahan2017communication,gu2021fast,wang2020tackling}.
Following existing literature \cite{li2020federated,mcmahan2017communication,gu2021fast,wang2020tackling}, we base our simulations on SVHN \cite{netzer2011reading}, CIFAR-10 \cite{krizhevsky2009learning} and CINIC-10 \cite{darlow2018cinic}.
All of them include $10$ classes of images of different categories.
For data heterogeneity, we partition all datasets and assign data samples to clients according to a Dirichlet distribution parameterized by $\alpha$ \cite{hsu2019measuring}.
In particular, $\alpha=0.1$ in Table.~\ref{tbl: exp main text}.
A smaller $\alpha$ entails a more non-\iid local data distribution and vice versa.
Each client holds the same data volume;  %
the exact data volume may be dataset-dependent.

\noindent{\bf Federated learning system.}
We consider %
$m=100$ clients, 
wherein clients continue to compute locally albeit the failures of unreliable communication uplinks. 
However, only clients with active links are allowed to submit their local updates. %
We use three customized convolutional neural networks 
for three datasets, respectively.
Next,
we introduce our construction of $p_i^t$'s,
which is then adopted to base the illustrations of unreliable patterns.

\noindent{\bf The construction of $p_i^t$'s.}
We define
\begin{align}
p_i^t \triangleq p_i \cdot \qth{(1-\gamma) + \gamma \cdot \epsilon^t},
\label{eq: definition pit}
\end{align}
where $p_i\in (0,1)$ is the time-invariant base probability,
$\gamma\in [0, 1]$ is time-invariant and is used to control the variations of $p_i^t$, 
and $\epsilon^t$ is time-dependent. 
Detailed specifications are forthcoming.

\begin{itemize}[leftmargin=*]
\item 
{\em Construction of $p_i$}.
The time-invariant base probability 
$p_i$ is jointly determined by the local data distribution %
and a random variable $R$,
which follows
a $\mathsf{lognormal}(\mu_0,\sigma_0^2)$ distribution.
Define the number of classes in a dataset as $C$, 
the class distribution at a client $i$ as $\bm{\nu}_i$ for $i\in [m]$.
Since the local datasets are partitioned according to $\mathsf{Dirichlet}(\alpha)$,
we have $\bm{\nu}_i \sim \mathsf{Dirichlet}(\alpha)$.
Sample $R$ from $\mathsf{lognormal}(\mu_0,\sigma_0^2)$ for $C$ times to obtain a positive vector $\bm{r}^\prime \in\reals^C$.
Normalize $\bm{r}^\prime$ by dividing its $l_1$ norm
and get $\bm{r} \triangleq \bm{r}^\prime / \|\bm{r}^\prime\|_1$. 
Finally, $p_i = \iprod{\bm{r}}{\bm{\nu}_i}$. 
Intuitively, $\bm{r}$ is used to quantify the unbalanced contribution of different classes. It is easy to see that for any fixed $\mu_0$, a larger $\sigma_0$ leads to a more heterogeneous contribution distribution.
We set
$\mu_0=0~\text{and}~\sigma_0 = 10$ in Table~\ref{tbl: exp main text}.
By definition, $p_i$
is a valid probability because 
\begin{align*}
0=\iprod{\bm{0}}{\bm{\nu}_i}
<
\iprod{\bm{r}}{\bm{\nu}_i} 
&\overset{(a)}{\le}
\iprod{\bm{r}}{\Indc} =1,
\end{align*}
where $\Indc$ is an all-one vector, 
$(a)$ holds because each element in $\bm{\nu}_i$ is no greater than 1,
and $p_i$ is strictly element-wise positive. 

\item
{\em Construction of $\epsilon^t$}.
\cite[Figure 5]{bonawitz2019towards} indicates that the number of participants,
\ie, clients with active communication uplinks, 
depends on time and acts like a sine curve.
Inspired by this, we introduce a time-varying noise
$\epsilon^t = \sin\qth{\pth{2 \pi / P }\cdot t}$, 
where $P=40$ defines the period and $t$ is the current round index. 
This is a similar setup as the {\em Home Device} unreliable communication scheme in \cite{ribero2022federated}.

\item 
{\em Choice of $\gamma$}.
By definition,
$\gamma$ in~\eqref{eq: definition pit} governs how severe the fluctuations of the sine curve in $p_i^t$'s are.
Given a fixed set of $p_i$'s, 
$\gamma$ determines both the lower and upper bounds of $p_i^t$'s.

\end{itemize}

\begin{figure}[!tb]
\centering
\begin{subfigure}[b]{\columnwidth}
\includegraphics[width=\columnwidth]{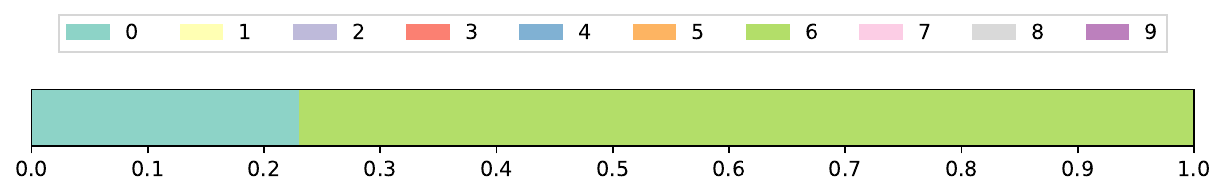}
\resizebox{\columnwidth}{!}{
\begin{tabular}{cccccccccc}
    \toprule
     \cellcolor[rgb]{0.5529411764705883, 0.8274509803921568, 0.7803921568627451} {\color{black} \small $2.30\times 10^{-1}$}   &
     \cellcolor[rgb]{1.0, 1.0, 0.7019607843137254} {\color{black} \small $4.50\times 10^{-11}$}   &
     \cellcolor[rgb]{0.7450980392156863, 0.7294117647058823, 0.8549019607843137} {\color{black} \small $1.03\times 10^{-10}$}   &
     \cellcolor[rgb]{0.984313725490196, 0.5019607843137255, 0.4470588235294118} {\color{black} \small $4.45\times 10^{-13}$}   &
     \cellcolor[rgb]{0.5019607843137255, 0.6941176470588235, 0.8274509803921568} {\color{black} \small $1.17\times 10^{-4}$}   &
     \cellcolor[rgb]{0.9921568627450981, 0.7058823529411765, 0.3843137254901961} {\color{black} \small $2.05\times 10^{-18}$}   &
     \cellcolor[rgb]{0.7019607843137254, 0.8705882352941177, 0.4117647058823529} {\color{black} \small $7.69\times 10^{-1}$}   &
     \cellcolor[rgb]{0.9882352941176471, 0.803921568627451, 0.8980392156862745} {\color{black} \small $1.01\times 10^{-11}$}   &
     \cellcolor[rgb]{0.8509803921568627, 0.8509803921568627, 0.8509803921568627} {\color{black} \small $4.94\times 10^{-7}$}   &
     \cellcolor[rgb]{0.7372549019607844, 0.5019607843137255, 0.7411764705882353} {\color{black} \small $1.68\times 10^{-9}$}\\
     \bottomrule
\end{tabular}
}    
\caption{An example of generated $\bm{r}$'s based on $\mathsf{lognormal}(0,10^2)$ distribution and normalization described above.
Each color corresponds to one class.
The first row visualizes the proportions of each class.
The second row presents the exact numbers (rounded up to 2 decimals). }
\label{fig: contribution plots}
\end{subfigure}
\begin{subfigure}[b]{\columnwidth}
\includegraphics[width=\columnwidth]{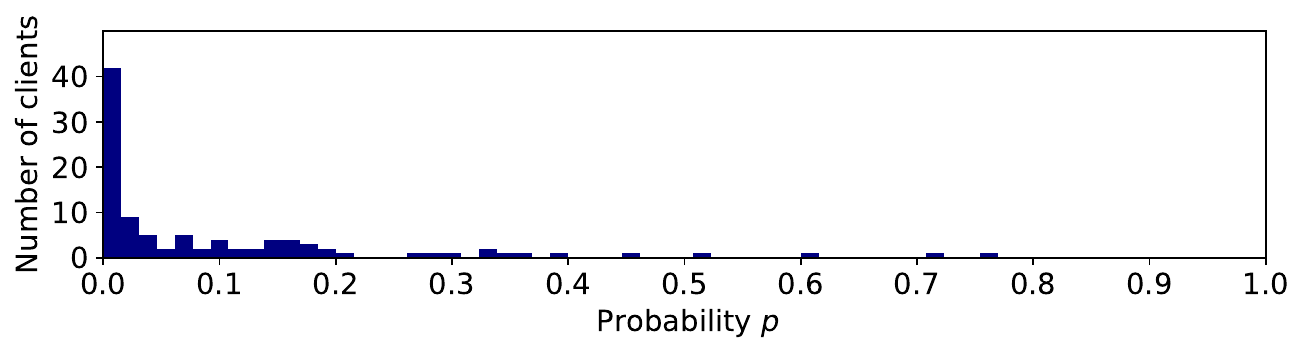}   
\caption{
Histograms of the constructed $p_i$'s under $R\sim\mathsf{lognormal}(0,10^2)$ and $\bm{\nu}_i\sim\mathsf{Dirichlet}(0.1)$ with $100$ clients and $\delta = 0$.
}
\label{fig: generated probabilities}
\end{subfigure}
\caption{\small The construction of $p_i$'s.}
\label{fig: pit constructions}
\end{figure}

~\prettyref{fig: contribution plots} presents an example of generated $\bm{r}$ drawn from a $\mathsf{lognormal}(0,10^2)$, wherein 
class $0$ and class $6$ dominate the entire distribution.
Intuitively,
if a client $i$ holds most of its images 
from classes other than $0$ or $6$,
the generated $p_i$ might be small and thus close to $0$,
possibly resulting in the client not appearing during training rounds in simulations.
See~\prettyref{fig: generated probabilities} for details.
To obtain meaningful results, we clip $p_i \gets \max\sth{\delta, p_i}$, where $\delta$ is a cutting-off parameter to ensure a lower bound on $p_i$.
In Table~\ref{tbl: exp main text}, $\delta = 0.02$.
Notably,
$\delta$ leads to the lower bound of $p_i^t$ being $\delta \cdot \pth{1 - 2\gamma}$.
Now,
we are ready to present unreliable schemes.

\noindent{\bf Unreliable schemes.}
In addition to a similar unreliable time-invariant communication setup as in \cite{wang2023lightweight} for fair competition,
we study a more challenging scenario where $p_i^t$'s change over time.
Specifically, we evaluate FedPBC and a set of baseline algorithms on the following schemes:
\begin{figure}[!tb]
\begin{subfigure}[b]{\textwidth}
\centering
\resizebox{.8\textwidth}{!}{
\tikzset{every picture/.style={line width=0.75pt, font=\large}} %
\begin{tikzpicture}[x=0.75pt,y=0.75pt,yscale=-1,xscale=1]

\draw    (110,195.5) -- (110,80.62) ;
\draw [shift={(110,77.62)}, rotate = 90] [fill={rgb, 255:red, 0; green, 0; blue, 0 }  ][line width=0.08]  [draw opacity=0] (7.14,-3.43) -- (0,0) -- (7.14,3.43) -- (4.74,0) -- cycle    ;
\draw    (110,195.5) -- (571.13,195.5) ;
\draw [shift={(574.13,195.5)}, rotate = 180] [fill={rgb, 255:red, 0; green, 0; blue, 0 }  ][line width=0.08]  [draw opacity=0] (7.14,-3.43) -- (0,0) -- (7.14,3.43) -- (4.74,0) -- cycle    ;
\draw   (157.5,135.62) -- (241.63,135.62) -- (241.63,195.5) -- (157.5,195.5) -- cycle ;
\draw   (273.5,135.62) -- (357.63,135.62) -- (357.63,195.5) -- (273.5,195.5) -- cycle ;
\draw   (390,135.62) -- (474.13,135.62) -- (474.13,195.5) -- (390,195.5) -- cycle ;
\draw    (392.75,167.06) -- (409.63,167.06) ;
\draw [shift={(389.75,167.06)}, rotate = 0] [fill={rgb, 255:red, 0; green, 0; blue, 0 }  ][line width=0.08]  [draw opacity=0] (5.36,-2.57) -- (0,0) -- (5.36,2.57) -- (3.56,0) -- cycle    ;
\draw    (470.63,167.56) -- (454.63,167.56) ;
\draw [shift={(473.63,167.56)}, rotate = 180] [fill={rgb, 255:red, 0; green, 0; blue, 0 }  ][line width=0.08]  [draw opacity=0] (5.36,-2.57) -- (0,0) -- (5.36,2.57) -- (3.56,0) -- cycle    ;

\draw    (160.75,165.56) -- (177.63,165.56) ;
\draw [shift={(157.75,165.56)}, rotate = 0] [fill={rgb, 255:red, 0; green, 0; blue, 0 }  ][line width=0.08]  [draw opacity=0] (5.36,-2.57) -- (0,0) -- (5.36,2.57) -- (3.56,0) -- cycle    ;
\draw    (238.63,166.06) -- (222.63,166.06) ;
\draw [shift={(241.63,166.06)}, rotate = 180] [fill={rgb, 255:red, 0; green, 0; blue, 0 }  ][line width=0.08]  [draw opacity=0] (5.36,-2.57) -- (0,0) -- (5.36,2.57) -- (3.56,0) -- cycle    ;

\draw    (277.25,167.06) -- (294.13,167.06) ;
\draw [shift={(274.25,167.06)}, rotate = 0] [fill={rgb, 255:red, 0; green, 0; blue, 0 }  ][line width=0.08]  [draw opacity=0] (5.36,-2.57) -- (0,0) -- (5.36,2.57) -- (3.56,0) -- cycle    ;
\draw    (355.13,167.56) -- (339.13,167.56) ;
\draw [shift={(358.13,167.56)}, rotate = 180] [fill={rgb, 255:red, 0; green, 0; blue, 0 }  ][line width=0.08]  [draw opacity=0] (5.36,-2.57) -- (0,0) -- (5.36,2.57) -- (3.56,0) -- cycle    ;

\draw  [dash pattern={on 0.84pt off 2.51pt}]  (247.25,166.56) -- (268.63,166.56) ;
\draw [shift={(271.63,166.56)}, rotate = 180] [fill={rgb, 255:red, 0; green, 0; blue, 0 }  ][line width=0.08]  [draw opacity=0] (5.36,-2.57) -- (0,0) -- (5.36,2.57) -- (3.56,0) -- cycle    ;
\draw [shift={(244.25,166.56)}, rotate = 0] [fill={rgb, 255:red, 0; green, 0; blue, 0 }  ][line width=0.08]  [draw opacity=0] (5.36,-2.57) -- (0,0) -- (5.36,2.57) -- (3.56,0) -- cycle    ;

\draw  [dash pattern={on 0.84pt off 2.51pt}]  (361.25,167.56) -- (386.63,167.56) ;
\draw [shift={(389.63,167.56)}, rotate = 180] [fill={rgb, 255:red, 0; green, 0; blue, 0 }  ][line width=0.08]  [draw opacity=0] (5.36,-2.57) -- (0,0) -- (5.36,2.57) -- (3.56,0) -- cycle    ;
\draw [shift={(358.25,167.56)}, rotate = 0] [fill={rgb, 255:red, 0; green, 0; blue, 0 }  ][line width=0.08]  [draw opacity=0] (5.36,-2.57) -- (0,0) -- (5.36,2.57) -- (3.56,0) -- cycle    ;
\draw    (388.65,112.73) .. controls (357.14,134.38) and (369.49,146.49) .. (371.94,167.56) ;
\draw [shift={(371.94,167.56)}, rotate = 263.37] [color={rgb, 255:red, 0; green, 0; blue, 0 }  ][line width=0.75]    (0,3.35) -- (0,-3.35)   ;
\draw [shift={(390.63,111.39)}, rotate = 146.5] [fill={rgb, 255:red, 0; green, 0; blue, 0 }  ][line width=0.08]  [draw opacity=0] (7.2,-1.8) -- (0,0) -- (7.2,1.8) -- cycle    ;
\draw [line width=0.75]    (157.5,195.5) -- (157.5,203.61) ;
\draw [line width=0.75]    (273.5,195.5) -- (273.5,203.61) ;
\draw    (160.75,201.06) -- (177.63,201.06) ;
\draw [shift={(157.75,201.06)}, rotate = 0] [fill={rgb, 255:red, 0; green, 0; blue, 0 }  ][line width=0.08]  [draw opacity=0] (5.36,-2.57) -- (0,0) -- (5.36,2.57) -- (3.56,0) -- cycle    ;
\draw    (269.63,201.56) -- (253.63,201.56) ;
\draw [shift={(272.63,201.56)}, rotate = 180] [fill={rgb, 255:red, 0; green, 0; blue, 0 }  ][line width=0.08]  [draw opacity=0] (5.36,-2.57) -- (0,0) -- (5.36,2.57) -- (3.56,0) -- cycle    ;
\draw [line width=0.75]    (390,195.5) -- (390,203.61) ;
\draw    (277.25,201.06) -- (294.13,201.06) ;
\draw [shift={(274.25,201.06)}, rotate = 0] [fill={rgb, 255:red, 0; green, 0; blue, 0 }  ][line width=0.08]  [draw opacity=0] (5.36,-2.57) -- (0,0) -- (5.36,2.57) -- (3.56,0) -- cycle    ;
\draw    (386.13,201.56) -- (370.13,201.56) ;
\draw [shift={(389.13,201.56)}, rotate = 180] [fill={rgb, 255:red, 0; green, 0; blue, 0 }  ][line width=0.08]  [draw opacity=0] (5.36,-2.57) -- (0,0) -- (5.36,2.57) -- (3.56,0) -- cycle    ;
\draw  [dash pattern={on 0.75pt off 0.75pt}]  (119.25,164.06) -- (122.25,164.06) .. controls (123.92,162.39) and (125.58,162.39) .. (127.25,164.06) .. controls (128.92,165.73) and (130.58,165.73) .. (132.25,164.06) .. controls (133.92,162.39) and (135.58,162.39) .. (137.25,164.06) .. controls (138.92,165.73) and (140.58,165.73) .. (142.25,164.06) -- (145.75,164.06) -- (148.75,164.06)(119.25,167.06) -- (122.25,167.06) .. controls (123.92,165.39) and (125.58,165.39) .. (127.25,167.06) .. controls (128.92,168.73) and (130.58,168.73) .. (132.25,167.06) .. controls (133.92,165.39) and (135.58,165.39) .. (137.25,167.06) .. controls (138.92,168.73) and (140.58,168.73) .. (142.25,167.06) -- (145.75,167.06) -- (148.75,167.06) ;
\draw [shift={(157.75,165.56)}, rotate = 180] [fill={rgb, 255:red, 0; green, 0; blue, 0 }  ][line width=0.08]  [draw opacity=0] (5.36,-2.57) -- (0,0) -- (5.36,2.57) -- (3.56,0) -- cycle    ;
\draw [shift={(110.25,165.56)}, rotate = 0] [fill={rgb, 255:red, 0; green, 0; blue, 0 }  ][line width=0.08]  [draw opacity=0] (5.36,-2.57) -- (0,0) -- (5.36,2.57) -- (3.56,0) -- cycle    ;
\draw    (174.93,81.71) .. controls (139.08,85.93) and (113.66,144.37) .. (133.13,163.03) ;
\draw [shift={(177.13,81.53)}, rotate = 176.91] [fill={rgb, 255:red, 0; green, 0; blue, 0 }  ][line width=0.08]  [draw opacity=0] (7.2,-1.8) -- (0,0) -- (7.2,1.8) -- cycle    ;
\draw    (489.51,147.13) .. controls (469.8,155.98) and (444.7,138.12) .. (432.63,154.53) ;
\draw [shift={(491.63,146.53)}, rotate = 162] [fill={rgb, 255:red, 0; green, 0; blue, 0 }  ][line width=0.08]  [draw opacity=0] (7.2,-1.8) -- (0,0) -- (7.2,1.8) -- cycle    ;
\draw   (158.47,214.6) .. controls (158.47,219.27) and (160.8,221.6) .. (165.47,221.6) -- (205.54,221.6) .. controls (212.21,221.6) and (215.54,223.93) .. (215.54,228.6) .. controls (215.54,223.93) and (218.87,221.6) .. (225.54,221.6)(222.54,221.6) -- (265.62,221.6) .. controls (270.29,221.6) and (272.62,219.27) .. (272.62,214.6) ;
\draw   (361.27,214.6) .. controls (361.27,219.27) and (363.6,221.6) .. (368.27,221.6) -- (408.34,221.6) .. controls (415.01,221.6) and (418.34,223.93) .. (418.34,228.6) .. controls (418.34,223.93) and (421.67,221.6) .. (428.34,221.6)(425.34,221.6) -- (468.42,221.6) .. controls (473.09,221.6) and (475.42,219.27) .. (475.42,214.6) ;

\draw (30,75) node [anchor=north west][inner sep=0.75pt]   [align=left] {\bf \em Link status};
\draw (63,129.5) node [anchor=north west][inner sep=0.75pt]   [align=left] {\it active};
\draw (55,183.5) node [anchor=north west][inner sep=0.75pt]   [align=left] {\it inactive};
\draw (489.5,199.5) node [anchor=north west][inner sep=0.75pt]   [align=left] {\bf \em Global round $\displaystyle t$};
\draw (110,197.9) node [anchor=north west][inner sep=0.75pt]    {$0$};
\draw (180.5,155) node [anchor=north west][inner sep=0.75pt]   [align=left] {\it active\\\it period};
\draw (297,155) node [anchor=north west][inner sep=0.75pt]   [align=left] {\it active\\\it period};
\draw (412.5,155) node [anchor=north west][inner sep=0.75pt]   [align=left] {\it active\\\it period};
\draw (520,170) node [anchor=north west][inner sep=0.75pt]   [align=left] {$\bm{\cdots}$};
\draw (391,98) node [anchor=north west][inner sep=0.75pt]   [align=left] {{\it inactive period}~$\displaystyle \triangleq ( 1-p_{i}) \cdot \text{cycle length}$};
\draw (175,200) node [anchor=north west][inner sep=0.75pt]   [align=left] {cycle length};
\draw (292,200) node [anchor=north west][inner sep=0.75pt]   [align=left] {cycle length};
\draw (178.5,71.5) node [anchor=north west][inner sep=0.75pt]   [align=left] {{\it random offset}~$\displaystyle \sim \mathsf{Uniform}\left[0,( 1-p_{i}) \cdot \text{cycle length}\right]$};
\draw (497,140) node [anchor=north west][inner sep=0.75pt]   [align=left] {$\displaystyle p_{i} \cdot \text{cycle length}$};
\draw (195,228) node [anchor=north west][inner sep=0.75pt]   [align=left] {\blue ({\bf fixed})};
\draw (395,228) node [anchor=north west][inner sep=0.75pt]   [align=left] {\blue ({\bf fixed})};

\end{tikzpicture}
}
\caption{\small
An illustration of cyclic {\em without} periodic reset,
where the communication link turns on and off in a cyclical fashion.
The length of a cycle is a predefined parameter.
Before a link becomes active for the first time,
it will remain off for a period of time, whose length is sampled from $\mathsf{Uniform} \qth{0, \pth{1 - p_i} \cdot \text{cycle length}}$.
After the initial stage,
the link will alternatively be in the active state with a fixed duration of the active period $\pth{p_i\cdot\text{cycle length}}$
or in the inactive state with a fixed duration of the inactive period $\qth{\pth{1 - p_i} \cdot \text{cycle length}}$.
In other words, 
the duration of the interval between two consecutive link switch-ons
is always fixed in length. 
}
\label{fig: cyclic without visualization}
\end{subfigure}
\vskip .5\baselineskip
\begin{subfigure}[b]{\textwidth}
\centering
\resizebox{.8\textwidth}{!}{
\tikzset{every picture/.style={line width=0.75pt,font=\large}} %

\begin{tikzpicture}[x=0.75pt,y=0.75pt,yscale=-1,xscale=1]

\draw    (110,195.5) -- (110,80.62) ;
\draw [shift={(110,77.62)}, rotate = 90] [fill={rgb, 255:red, 0; green, 0; blue, 0 }  ][line width=0.08]  [draw opacity=0] (7.14,-3.43) -- (0,0) -- (7.14,3.43) -- (4.74,0) -- cycle    ;
\draw    (110,195.5) -- (571.13,195.5) ;
\draw [shift={(574.13,195.5)}, rotate = 180] [fill={rgb, 255:red, 0; green, 0; blue, 0 }  ][line width=0.08]  [draw opacity=0] (7.14,-3.43) -- (0,0) -- (7.14,3.43) -- (4.74,0) -- cycle    ;
\draw   (133.5,135.62) -- (217.63,135.62) -- (217.63,195.5) -- (133.5,195.5) -- cycle ;
\draw    (136.75,165.56) -- (153.63,165.56) ;
\draw [shift={(133.75,165.56)}, rotate = 0] [fill={rgb, 255:red, 0; green, 0; blue, 0 }  ][line width=0.08]  [draw opacity=0] (5.36,-2.57) -- (0,0) -- (5.36,2.57) -- (3.56,0) -- cycle    ;
\draw    (214.63,166.06) -- (198.63,166.06) ;
\draw [shift={(217.63,166.06)}, rotate = 180] [fill={rgb, 255:red, 0; green, 0; blue, 0 }  ][line width=0.08]  [draw opacity=0] (5.36,-2.57) -- (0,0) -- (5.36,2.57) -- (3.56,0) -- cycle    ;

\draw   (258.5,135.62) -- (342.63,135.62) -- (342.63,195.5) -- (258.5,195.5) -- cycle ;
\draw    (262.25,167.06) -- (279.13,167.06) ;
\draw [shift={(259.25,167.06)}, rotate = 0] [fill={rgb, 255:red, 0; green, 0; blue, 0 }  ][line width=0.08]  [draw opacity=0] (5.36,-2.57) -- (0,0) -- (5.36,2.57) -- (3.56,0) -- cycle    ;
\draw    (340.13,167.56) -- (324.13,167.56) ;
\draw [shift={(343.13,167.56)}, rotate = 180] [fill={rgb, 255:red, 0; green, 0; blue, 0 }  ][line width=0.08]  [draw opacity=0] (5.36,-2.57) -- (0,0) -- (5.36,2.57) -- (3.56,0) -- cycle    ;

\draw   (368,135.62) -- (452.13,135.62) -- (452.13,195.5) -- (368,195.5) -- cycle ;
\draw    (370.75,167.06) -- (387.63,167.06) ;
\draw [shift={(367.75,167.06)}, rotate = 0] [fill={rgb, 255:red, 0; green, 0; blue, 0 }  ][line width=0.08]  [draw opacity=0] (5.36,-2.57) -- (0,0) -- (5.36,2.57) -- (3.56,0) -- cycle    ;
\draw    (448.63,167.56) -- (432.63,167.56) ;
\draw [shift={(451.63,167.56)}, rotate = 180] [fill={rgb, 255:red, 0; green, 0; blue, 0 }  ][line width=0.08]  [draw opacity=0] (5.36,-2.57) -- (0,0) -- (5.36,2.57) -- (3.56,0) -- cycle    ;

\draw [line width=0.75]    (110.05,195.5) -- (110.05,203.61) ;
\draw [line width=0.75]    (226.05,195.5) -- (226.05,203.61) ;
\draw    (113.3,201.06) -- (130.18,201.06) ;
\draw [shift={(110.3,201.06)}, rotate = 0] [fill={rgb, 255:red, 0; green, 0; blue, 0 }  ][line width=0.08]  [draw opacity=0] (5.36,-2.57) -- (0,0) -- (5.36,2.57) -- (3.56,0) -- cycle    ;
\draw    (222.18,201.56) -- (206.18,201.56) ;
\draw [shift={(225.18,201.56)}, rotate = 180] [fill={rgb, 255:red, 0; green, 0; blue, 0 }  ][line width=0.08]  [draw opacity=0] (5.36,-2.57) -- (0,0) -- (5.36,2.57) -- (3.56,0) -- cycle    ;
\draw [line width=0.75]    (342.55,195.5) -- (342.55,203.61) ;
\draw    (229.8,201.06) -- (246.68,201.06) ;
\draw [shift={(226.8,201.06)}, rotate = 0] [fill={rgb, 255:red, 0; green, 0; blue, 0 }  ][line width=0.08]  [draw opacity=0] (5.36,-2.57) -- (0,0) -- (5.36,2.57) -- (3.56,0) -- cycle    ;
\draw    (338.68,201.56) -- (322.68,201.56) ;
\draw [shift={(341.68,201.56)}, rotate = 180] [fill={rgb, 255:red, 0; green, 0; blue, 0 }  ][line width=0.08]  [draw opacity=0] (5.36,-2.57) -- (0,0) -- (5.36,2.57) -- (3.56,0) -- cycle    ;
\draw  [dash pattern={on 0.75pt off 0.75pt}]  (118.93,164.06) -- (121.93,164.06) -- (121.75,164.06) -- (124.75,164.06)(118.93,167.06) -- (121.93,167.06) -- (121.75,167.06) -- (124.75,167.06) ;
\draw [shift={(133.75,165.56)}, rotate = 180] [fill={rgb, 255:red, 0; green, 0; blue, 0 }  ][line width=0.08]  [draw opacity=0] (5.36,-2.57) -- (0,0) -- (5.36,2.57) -- (3.56,0) -- cycle    ;
\draw [shift={(109.93,165.56)}, rotate = 0] [fill={rgb, 255:red, 0; green, 0; blue, 0 }  ][line width=0.08]  [draw opacity=0] (5.36,-2.57) -- (0,0) -- (5.36,2.57) -- (3.56,0) -- cycle    ;
\draw    (163.32,116.62) .. controls (135.38,121.64) and (104.57,149.41) .. (124.13,163.03) ;
\draw [shift={(165.46,116.28)}, rotate = 172.14] [fill={rgb, 255:red, 0; green, 0; blue, 0 }  ][line width=0.08]  [draw opacity=0] (7.2,-1.8) -- (0,0) -- (7.2,1.8) -- cycle    ;
\draw    (489.51,147.13) .. controls (469.8,151.98) and (444.7,138.12) .. (432.63,154.53) ;
\draw [shift={(491.63,146.53)}, rotate = 162] [fill={rgb, 255:red, 0; green, 0; blue, 0 }  ][line width=0.08]  [draw opacity=0] (7.2,-1.8) -- (0,0) -- (7.2,1.8) -- cycle    ;
\draw  [dash pattern={on 0.75pt off 0.75pt}]  (235.1,164.97) -- (238.1,165.05) .. controls (239.81,163.42) and (241.47,163.46) .. (243.1,165.17) -- (247.29,165.27) -- (250.29,165.34)(235.03,167.97) -- (238.03,168.05) .. controls (239.74,166.42) and (241.4,166.46) .. (243.03,168.17) -- (247.22,168.27) -- (250.22,168.34) ;
\draw [shift={(259.25,167.06)}, rotate = 181.38] [fill={rgb, 255:red, 0; green, 0; blue, 0 }  ][line width=0.08]  [draw opacity=0] (5.36,-2.57) -- (0,0) -- (5.36,2.57) -- (3.56,0) -- cycle    ;
\draw [shift={(226.07,166.26)}, rotate = 1.38] [fill={rgb, 255:red, 0; green, 0; blue, 0 }  ][line width=0.08]  [draw opacity=0] (5.36,-2.57) -- (0,0) -- (5.36,2.57) -- (3.56,0) -- cycle    ;
\draw    (226.05,166.26) -- (226.05,195.5) ;
\draw    (242.22,120.36) .. controls (239.87,132.88) and (222.19,154.97) .. (242.66,166.66) ;
\draw [shift={(242.46,118.28)}, rotate = 91.91] [fill={rgb, 255:red, 0; green, 0; blue, 0 }  ][line width=0.08]  [draw opacity=0] (7.2,-1.8) -- (0,0) -- (7.2,1.8) -- cycle    ;
\draw  [dash pattern={on 0.75pt off 0.75pt}]  (352.13,166.06) -- (355.13,166.06) -- (355.04,166.06) -- (358.04,166.06)(352.13,169.06) -- (355.13,169.06) -- (355.04,169.06) -- (358.04,169.06) ;
\draw [shift={(367.04,167.56)}, rotate = 180] [fill={rgb, 255:red, 0; green, 0; blue, 0 }  ][line width=0.08]  [draw opacity=0] (5.36,-2.57) -- (0,0) -- (5.36,2.57) -- (3.56,0) -- cycle    ;
\draw [shift={(343.13,167.56)}, rotate = 0] [fill={rgb, 255:red, 0; green, 0; blue, 0 }  ][line width=0.08]  [draw opacity=0] (5.36,-2.57) -- (0,0) -- (5.36,2.57) -- (3.56,0) -- cycle    ;
\draw [line width=0.75]    (458.98,196.04) -- (458.98,204.15) ;
\draw    (346.23,201.6) -- (363.11,201.6) ;
\draw [shift={(343.23,201.6)}, rotate = 0] [fill={rgb, 255:red, 0; green, 0; blue, 0 }  ][line width=0.08]  [draw opacity=0] (5.36,-2.57) -- (0,0) -- (5.36,2.57) -- (3.56,0) -- cycle    ;
\draw    (455.11,202.1) -- (439.11,202.1) ;
\draw [shift={(458.11,202.1)}, rotate = 180] [fill={rgb, 255:red, 0; green, 0; blue, 0 }  ][line width=0.08]  [draw opacity=0] (5.36,-2.57) -- (0,0) -- (5.36,2.57) -- (3.56,0) -- cycle    ;
\draw    (321.04,114.16) .. controls (341.78,137.73) and (366.96,127.72) .. (354.62,164.42) ;
\draw [shift={(319.46,112.28)}, rotate = 51.07] [fill={rgb, 255:red, 0; green, 0; blue, 0 }  ][line width=0.08]  [draw opacity=0] (7.2,-1.8) -- (0,0) -- (7.2,1.8) -- cycle    ;
\draw   (130.63,218.93) .. controls (132.63,222.6) and (135.96,224.93) .. (140.63,224.94) -- (185.38,224.96) .. controls (192.05,224.97) and (195.38,227.3) .. (195.38,231.97) .. controls (195.38,227.3) and (198.71,224.97) .. (205.38,224.97)(202.38,224.97) -- (250.13,225) .. controls (254.8,225) and (260.13,222.67) .. (262.13,217) ;
\draw   (343.63,218.93) .. controls (343.63,223.6) and (345.96,225.93) .. (350.63,225.93) -- (387.63,225.93) .. controls (394.3,225.93) and (397.63,228.26) .. (397.63,232.93) .. controls (397.63,228.26) and (400.96,225.93) .. (407.63,225.93)(404.63,225.93) -- (444.63,225.93) .. controls (449.3,225.93) and (451.63,223.6) .. (451.63,218.93) ;
\draw (30,75) node [anchor=north west][inner sep=0.75pt]   [align=left] {\bf \em Link status};
\draw (63,129.5) node [anchor=north west][inner sep=0.75pt]   [align=left] {\it active};
\draw (55,183.5) node [anchor=north west][inner sep=0.75pt]   [align=left] {\it inactive};
\draw (489.5,199.5) node [anchor=north west][inner sep=0.75pt]   [align=left] {\bf \em Global round $\displaystyle t$};
\draw (109.17,205.9) node [anchor=north west][inner sep=0.75pt]    {$0$};
\draw (155,156) node [anchor=north west][inner sep=0.75pt]  [align=left] {\it active\\\it period};
\draw (280,157.5) node [anchor=north west][inner sep=0.75pt]  [align=left] {\it active\\\it period};
\draw (390,157.5) node [anchor=north west][inner sep=0.75pt]  [align=left] {\it active\\\it period};
\draw (128,201) node [anchor=north west][inner sep=0.75pt]  [align=left] {cycle length};
\draw (245,201) node [anchor=north west][inner sep=0.75pt]  [align=left] {cycle length};
\draw (497,140) node [anchor=north west][inner sep=0.75pt]  [align=left] {$\displaystyle p_{i} \cdot \text{cycle length}$};
\draw (157.14,235) node [anchor=north west][inner sep=0.75pt]   [align=left] {\blue ({\bf stochastic})};
\draw (135.5,92.5) node [anchor=north west][inner sep=0.75pt]  [align=left] {{\it random offset}~$\displaystyle \sim \mathsf{Uniform}\left[0,( 1-p_{i}) \cdot \text{cycle length}\right]$};
\draw (360,201) node [anchor=north west][inner sep=0.75pt]  [align=left] {cycle length};
\draw (365,235) node [anchor=north west][inner sep=0.75pt]   [align=left] {\blue ({\bf stochastic})};

\end{tikzpicture}
}
\caption{\small
An illustration of cyclic {\em with} periodic reset.
Similar to~\prettyref{fig: cyclic without visualization}, 
a link switches on and off in alternation.
The key difference is that
a random offset will be redrawn from the same uniform distribution at the beginning of each cycle.
The resampling procedure is called a reset,
which entails a stochastic length of the interval between two consecutive link switch-ons.
}
\label{fig: cyclic with visualization}
\end{subfigure}
\caption{\small Illustrations of the communication unreliable schemes evaluated in~\prettyref{sec: real world numerical}}
\label{fig: unreliable scheme}
\end{figure}
\vskip -.5\baselineskip

\begin{enumerate}[leftmargin=*]
\item 
{\bf Bernoulli.}
Client $i$ submits its local updates to the parameter server when the uplink becomes active with probability $p_i^t$.
The first two columns of Table~\ref{tbl: exp main text} demonstrate the results when the probabilities are {\em time-invariant} $p_i$'s and {\em time-varying} $p_i^t$'s, respectively.
When $p_i^t$ is time-invariant, we have $p_i^t = p_i$ for all $t \ge 0$,
where $p_i$ is the time-invariant base probability in~\eqref{eq: definition pit}.
In the latter,
$p_i^t$ is defined as in~\eqref{eq: definition pit}
and changes over time.
\item 
{\bf Markovian.}
The uplink connection probabilities $p_i^t$'s 
might be affected by external factors,  
leading to an unexpected shutdown after it is on
or,
conversely,
resuming fully operational after it is off.
Specifically,
the uplink availability is dictated by 
a Markov chain of two states ``ON'' and ``OFF'',
whose initial state is determined by a Bernoulli sampling.
Depending on whether 
the transition probabilities
change over time,
we have a homogeneous Markov chain 
(the third row of Table~\prettyref{tbl: exp main text})
or
a non-homogeneous Markov chain 
(the fourth row).
The detailed illustration of the transition probabilities is deferred to~\prettyref{app: exp}.

\item 
{\bf Cyclic.}
The communication uplink between the parameter server and the clients can have a cyclic pattern,
where the client has a fixed working schedule and joins the training diurnally or nocturnally \cite{pmlr-v202-cho23b,bonawitz2019towards}.
A random offset at the beginning of the whole process is used to simulate and reflect the initial shift due to each client's device heterogeneity \cite{wang2023lightweight}.
Please refer to~\prettyref{fig: cyclic without visualization} for details.
However, 
it is also possible that each client's schedule to start training varies each day,
which motivates us to devise the second scheme with periodic reset in~\prettyref{fig: cyclic with visualization}.
The key difference is that the random offset will be reset at the beginning of each cycle, not only at the first cycle.
Notice that the interval for a link to become active is now stochastic, rather than fixed.
\end{enumerate}

\begin{figure}[!tb]
\centering
\begin{subfigure}[b]{\textwidth}
\centering
\includegraphics[width=.8\textwidth]{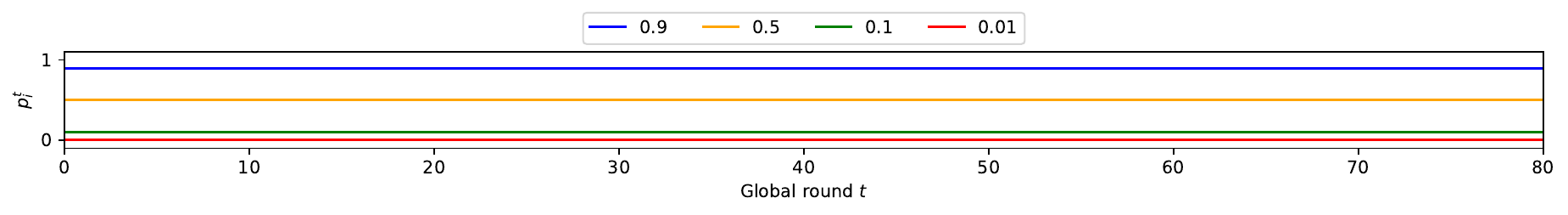}
\includegraphics[width=.8\textwidth]{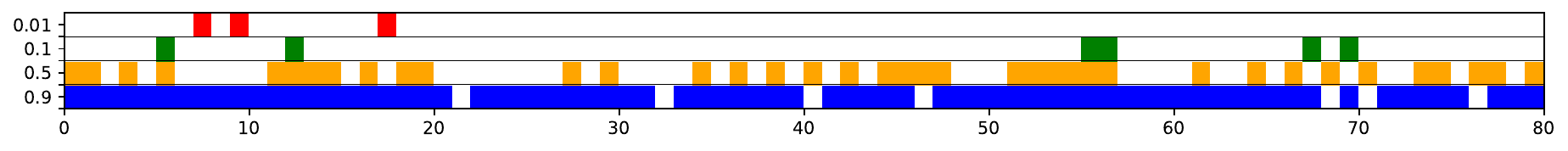}
\vskip -.5\baselineskip
\caption{
Bernoulli with {\em time-invariant} $p_i^t=p_i$'s in a total of $80$ global rounds.
The first row shows the trajectories of time-invariant $p_i^t=p_i$'s .
The second row shows the status of the uplink sampled from $\mathsf{Bernoulli}(p_i)$.
}
\label{fig: time-invariant bernoulli visualization}
\end{subfigure}
\begin{subfigure}[b]{\textwidth}
\centering
\includegraphics[width=.8\linewidth]{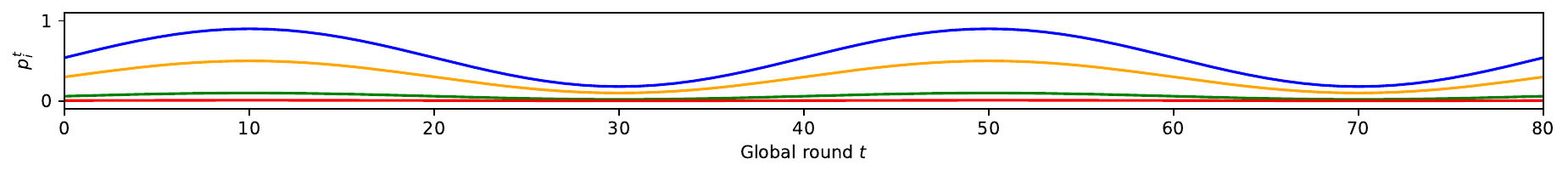}
\includegraphics[width=.8\linewidth]{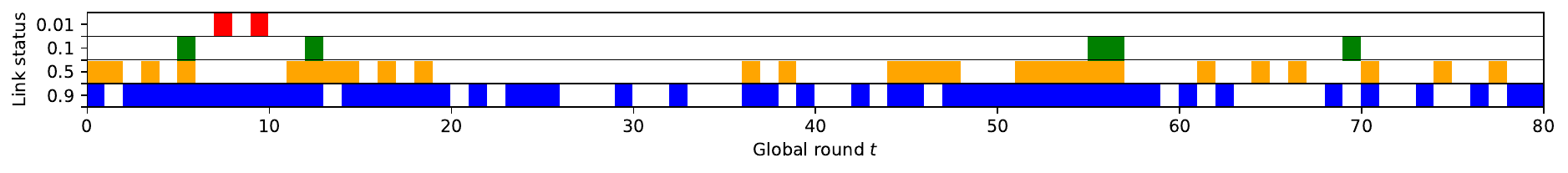}
\vskip -.5\baselineskip
\caption{
Bernoulli with {\em time-varying} $p_i^t$'s in a total of $80$ global rounds.
The first row shows the trajectories of time-varying $p_i^t$'s.
The second row shows the status of the uplink sampled from $\mathsf{Bernoulli}(p_i^t)$.
}
\label{fig: time-varying bernoulli visualization}
\end{subfigure}
\begin{subfigure}[b]{\textwidth}
\centering
\includegraphics[width=.8\linewidth]{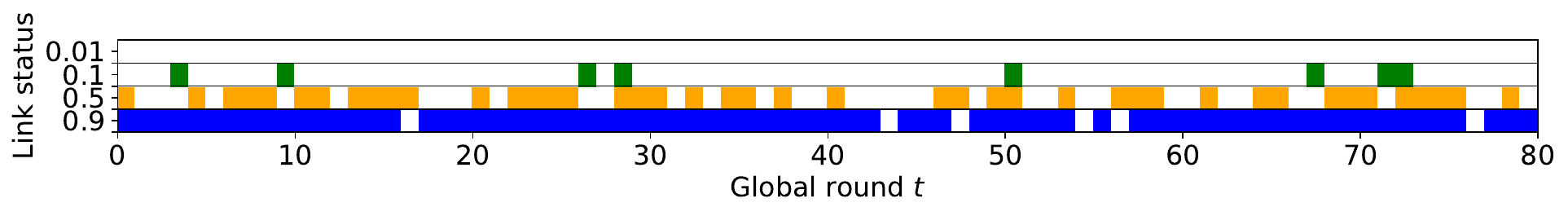}
\vskip -.5\baselineskip
\caption{
The status of the uplink under homogeneous Markovian in a total of $80$ global rounds.
}
\label{fig: time-invariant markovian visualization}
\end{subfigure}
\begin{subfigure}[b]{\textwidth}
\centering
\includegraphics[width=.8\linewidth]{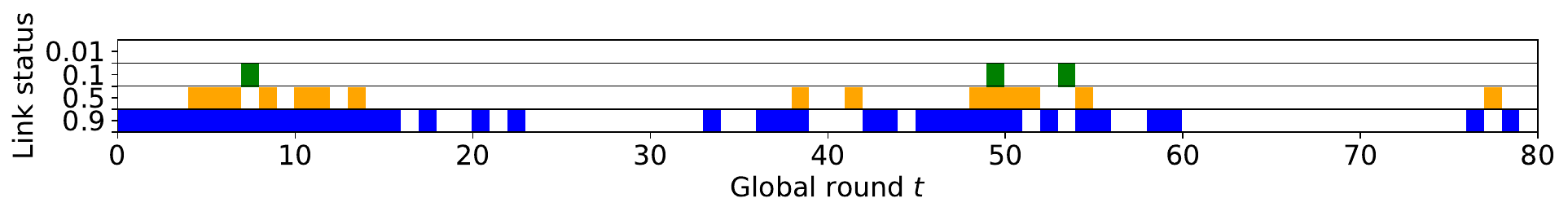}
\vskip -.5\baselineskip
\caption{
The status of the uplink under heterogeneous Markovian in a total of $80$ global rounds.
}
\label{fig: time-varying markovian visualization}
\end{subfigure}
\begin{subfigure}[b]{\textwidth}
\centering
\includegraphics[width=.8\linewidth]{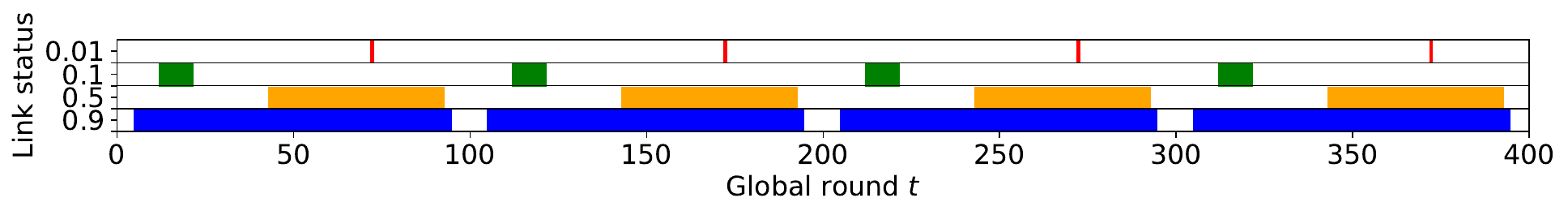}
\vskip -.5\baselineskip
\caption{
The status of the uplink under cyclic without periodic reset in a total of $400$ global rounds.
The cycle length is $100$.
}
\label{fig: time-invariant cyclic visualization}
\end{subfigure}
\begin{subfigure}[b]{\textwidth}
\centering
\includegraphics[width=.8\linewidth]{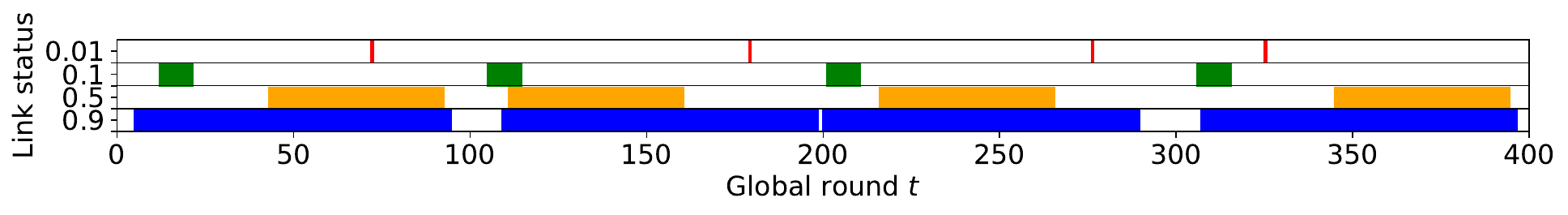}
\vskip -.5\baselineskip
\caption{
The status of the uplink under cyclic with periodic reset in a total of $400$ global rounds.
The cycle length is $100$.
}
\label{fig: time-varying cyclic visualization}
\end{subfigure}
\caption{\small
Exemplary trajectories of $p_i^t$'s and uplink status under different unreliable communication schemes.
Colored blocks indicate that an uplink is active in the given round.
We simulate the scenarios where $p_i \in \sth{0.01, 0.1, 0.5, 0.9}$.
The construction of $p_i^t$ based on $p_i$ can be found in~\prettyref{sec: real world numerical}.
}
\label{fig: computed realizations}
\end{figure}

\prettyref{fig: computed realizations} shows an example of uplink statuses under the unreliable communication schemes we evaluate.
It is observed that uplinks become less frequently active when probabilities change from time-invariant (\prettyref{fig: time-invariant bernoulli visualization}) to time-varying (\prettyref{fig: time-varying bernoulli visualization}).
In addition, 
the uplinks become even more sparsely active when the schemes move to Markovian in~\prettyref{fig: time-invariant markovian visualization} and \ref{fig: time-varying markovian visualization}.
On the other hand, 
the cyclic unreliable scheme exhibits a different pattern:
the uplinks in~\prettyref{fig: cyclic without visualization} become active and inactive in alternation
after an initial random offset.
Notice that 
the uplink's offline duration 
is always fixed.
In contrast,
the duration %
remains random in~\prettyref{fig: cyclic with visualization} due to a reset at the beginning of each cycle.

\noindent{\bf Baseline algorithms.}
We compare FedPBC with six baseline algorithms,
including
FedAvg \cite{mcmahan2017communication},
FedAvg {\em all},
FedAvg {\em known $p_i^t$'s} \cite{perazzone2022communication},
FedAU \cite{wang2023lightweight},
F3AST \cite{ribero2022federated},
and MIFA \cite{gu2021fast}.
Under FedAvg {\em all}, 
the parameter server averages all clients' local updates, 
wherein the contributions of clients with inactive communication links are deemed zeros.
FedAvg {\em known $p_i^t$'s} requires the time-varying $p_i^t$'s to be a known prior.
We defer the other algorithmic specific parameters to~\prettyref{app: exp}.

\noindent{\bf Results.}
Table~\ref{tbl: exp main text} presents the evaluation results.
In summary,
FedPBC outperforms all other baseline algorithms {\em not} aided by memory
on the SVHN and CINIC-10 datasets. 
In a rare instance, 
FedPBC is surpassed by FedAU on the CIFAR-10 dataset by a mere 0.2\% in test accuracy.
The rationale merits additional scrutiny.
Additionally, 
FedAvg trails behind FedPBC by a substantial margin of approximately 10\% in test accuracy, confirming its inherent bias. 

\FloatBarrier
\begin{table}[H]
\centering
\caption{\small 
The first round to reach a targeted test accuracy under Bernoulli with {\em time-varying} $p_i^t$'s over 3 random seeds.
We study the first round to reach $1/4$, $1/2$, $3/4$ and $1$ of the best test accuracy of each dataset in Table~\ref{tbl: exp main text},
which is rounded up to the nearest $10\%$ below for ease of presentation.
In addition,
we sample the mean of test accuracy every 150 global rounds to mitigate %
noisy progress. 
Some algorithms may never attain the targeted accuracy
due to their inferior performance,
where we use ''--'' as a placeholder.
For example,
the best test accuracy of FedAvg {\em all}
is 36.5\% under Bernoulli with {\em time-varying} $p_i^t$'s in Table~\ref{tbl: exp main text},
below both 3/4 and 1 of the best accuracy.
}
\label{tbl: slowdown main text}
\resizebox{.6\textwidth}{!}{
\begin{tabular}{c|c|c|c|c|c}
\toprule
\multirow{2}{*}{\bf Datasets}&
{\bf Quarters}&
{\bf 1/4}&
{\bf 1/2}&
{\bf 3/4}&
{\bf 1}
\\
\noalign{\vspace{.5mm}}
\cline{2-6}
\noalign{\vspace{.8mm}}
&
{\bf Test accuracy}&
{\bf $20\%$}&
{\bf $40\%$}&
{\bf $60\%$}&
{\bf $80\%$}
\\
\midrule
\multirow{7}{*}{SVHN}&
FedPBC ({ours})& 
150 &
300 &
450 &
1650 \\
& 
FedAvg & 
300 &
450 &
1050 &
-- 
\\
& 
FedAvg {\em all} & 
1950 &
-- &
-- &
--
\\& 
FedAU & 
300 &
300 &
750 &
3450
\\
& 
F3AST & %
450 &
750 &
1200 &
3600
\\
\noalign{\vspace{.5mm}}
\cline{2-6}
\noalign{\vspace{.5mm}}
& 
FedAvg {\em known} $p_i^t$'s 
& 
600 &
1050 &
1650 &
--
\\
& 
MIFA ({\em memory aided}) & 
300 &
600 &
1050&
-- \\
\midrule
\multicolumn{1}{c}{}&
{\bf Test accuracy}&
{\bf $15\%$}&
{\bf $30\%$}&
{\bf $45\%$}&
{\bf $60\%$} \\
\midrule
\multirow{7}{*}{CIFAR-10}&
FedPBC ({ours})& 
150 &
150 &
450 &
3300
\\& 
FedAvg & 
150 &
450 &
1050 &
9450
\\
& 
FedAvg {\em all} & 
150 &
1500 &
-- &
--
\\
& 
FedAU & 
150 &
300 &
750 &
3900
\\
& 
F3AST & %
150 &
300 &
1200 &
4800
\\
\noalign{\vspace{.5mm}}
\cline{2-6}
\noalign{\vspace{.5mm}}
& 
FedAvg {\em known} $p_i^t$'s 
& 
0 &
450 &
1800 &
4800
\\
& 
MIFA ({\em memory aided}) & 
150 &
150 &
600&
3600 \\
\midrule
\multicolumn{1}{c}{}&
{\bf Test accuracy}&
{\bf $10\%$}&
{\bf $20\%$}&
{\bf $30\%$}&
{\bf $40\%$} \\
\midrule
\multirow{7}{*}{CINIC-10}&
FedPBC ({ours})& 
\multirow{5}{*}{0} &
150 &
300 &
900
\\
& 
FedAvg & 
 &
150 &
1050 &
6450
\\
& 
FedAvg {\em all} & 
 &
600 &
-- &
--
\\
& 
FedAU & 
 &
150 &
300 &
2700
\\
& 
F3AST & %
 &
300 &
1200 &
3000 
\\
\noalign{\vspace{.5mm}}
\cline{2-6}
\noalign{\vspace{.5mm}}
& 
FedAvg {\em known} $p_i^t$'s & 
\multirow{2}{*}{0}&
300 &
1050 &
2850
\\
& 
MIFA ({\em memory aided}) & 
 &
150 &
900&
2700 \\
\bottomrule
\end{tabular}}
\end{table}

It turns out that MIFA,
aided by $100$ units of old local gradients,
does not always achieve the best performance.
We conjecture it to the old gradients induced by a lower participation rate.
\prettyref{fig: generated probabilities} shows that most probabilities fall below $0.1$ under our construction of $p_i$' s, which means that an uplink could be inactive for a long time before waking up again.
Although clients in FedPBC start in each global round from its own staled local model,
the expected staleness is upper bounded (see \prettyref{prop: staleness term}).
It is not surprising that F3AST acts inferior to FedPBC.
At a high level,
F3AST caps $\calA^t$ to a few representative clients for local optimization, %
excluding the rest of the clients within $\calA^t$. 
FedPBC surpasses FedAU in all scenarios in terms of train accuracy.
Although FedAU develops an online average method to estimate the underlying connection probabilities, %
it cannot tolerate complex dynamics. 
This can be observed in the performance degradation when switching from cyclic {\em without} periodic restart to cyclic {\em with} periodic restart. 
In the former, the uplinks are activated alternately with a fixed interval after the initial random offset, whereas in the latter, they are switched on stochastically, making it much more challenging. 
In the case of time-invariant $p_i$'s, the outperformance of our 
FedPBC may stem from its utilization of true gradient trajectories to account for inactivities. This approach may result in better compensation than the online estimate used in FedAU.
Though FedAvg with {\em known} probability uses the ground truth $1/p_i^t$ to mimic the empirical length of the uplink active interval,
as pointed out in \cite{wang2023lightweight},
the empirical length can unfortunately deviate far from the ground truth $1/p_i^t$.

To complement the numerical results in the main section,
we also study the impact of different system-design parameters,
including $\alpha,~\gamma,~\delta,~\sigma_0$,
on learning performance.
The results are deferred to~\prettyref{app: exp}.

\noindent{\bf Staleness.} 
Table~\ref{tbl: slowdown main text} demonstrates the first round to reach a targeted test accuracy under Benoulli with {\em time-varying} $p_i^t$'s.
Specifically,
we study the round to reach the four quarters of the best test accuracy,
which is rounded to the nearest 10\% below for a neat presentation.
It is readily seen that 
FedPBC attains 
a similar round to reach
1/4 and 1/2 of the best test accuracy
as either FedAU or MIFA.
When it is beyond 3/4 of the best accuracy,
FedPBC in fact becomes the fastest algorithm.
Hence,
we empirically conclude that the staleness in FedPBC is mild
and confirms its practicality.

\section{Proofs of Selected Results}
\label{sec: selected proofs}
In this section, we present proofs of Lemma~\ref{lemma: descent lemma} and \ref{lemma: consensus}.
Proposition~\ref{proposition: average gradient to global gradient} is illustrated first as an intermediate result to assist in the proofs.
\begin{proposition}
\label{proposition: average gradient to global gradient}
For any $t\in[T-1]$, it holds that 
\begin{align}
\nonumber
\frac{1}{m}\sum_{i=1}^m\norm{\nabla F_i(\x_i^t)}^2 \le \frac{3L^2}{m} \sum_{i=1}^m \norm{\x_i^t - \bar{\x}^t}^2 + 3\zeta^2 \\
\label{eq: gradient heterogeneity}
+ 3\pth{\beta^2 + 1}\norm{\nabla F(\bar{\x}^t)}^2 . 
\end{align}
\end{proposition}
Inequality~\eqref{eq: gradient heterogeneity} can be shown by Jensen's inequality, where we plug in Assumptions~\ref{ass: 2 smmothness} and~\ref{ass: bounded similarity}.
\begin{proof}[\bf Proof of Lemma \ref{lemma: descent lemma}]
By~\prettyref{ass: 2 smmothness},
we have 
\begin{align*}
&F(\bar{\x}^{t+1})  -  F(\bar{\x}^{t}) \le \iprod{\nabla F(\bar{\x}^{t})}{\bar{\x}^{t+1} - \bar{\x}^{t}} + \frac{L}{2}\norm{\bar{\x}^{t+1}- \bar{\x}^{t}}^2 \\
& = \iprod{\nabla F(\bar{\x}^{t})}{- \frac{\eta}{m} \bm{G}^{(t)} \bm{1}} +  \frac{L\eta^2}{2}\norm{\frac{1}{m} \bm{G}^{(t)} \bm{1}}^2.  
\end{align*}
Taking expectations with respect to the randomness in the mini-batches at $t$-th rounds, we have 
\begin{align*}
&\expect{F(\bar{\x}^{t+1})  -  F(\bar{\x}^{t}) \mid \calF^{t}} \\
&\le \expect{\iprod{\nabla F(\bar{\x}^{t})}{- \frac{\eta}{m} \bm{G}^{(t)} \bm{1}} +  \frac{L \eta^2}{2}\norm{\frac{1}{m} \bm{G}^{(t)} \bm{1}}^2 \mid \calF^{t}}. 
\end{align*}
For ease of notations, we abbreviate $\nabla \ell_i(\x_i^{\pth{t,k}})$ as $\nabla \ell_i^{\pth{t,k}}.$
\noindent{\em (a) Bounding
$\mathbb{E}[\iprod{\nabla f(\bar{\x}^{t})}{- \frac{\eta}{m} \nabla \bm{G}^{(t)} \bm{1}} \mid \calF^{t}]$.
}
\begin{align*}
&\expect{\iprod{\nabla F(\bar{\x}^{t})}{- \frac{\eta}{m} \bm{G}^{(t)} \bm{1}} \mid \calF^{t}} \\
&= - \frac{\eta}{m} \expect{\iprod{\nabla F(\bar{\x}^{t})}{\sum_{i=1}^m \sum_{k=0}^{s-1} \nabla \ell_i^{(t,k)}} \mid \calF^{t}}    \\
& = \underbrace{- \frac{s\eta}{m} \iprod{\nabla F(\bar{\x}^{t})}{\nabla \bm{F}^{(t)}\bm{1}}}_{(\rmA)} \\
&~~~+ \underbrace{\expect{\frac{\eta}{m} \iprod{\nabla F(\bar{\x}^{t})}{\sum_{i=1}^m s \nabla \ell_i^{(t,0)} - \sum_{k=0}^{s-1} \nabla \ell_i^{(t,k)} }\mid \calF^{t}}}_{(\rmB)}.  
\end{align*}

Term $(\rmA)$ can be bounded as 
\begin{align*}
&- s\eta\iprod{\nabla F(\bar{\x}^{t})}{ \frac{1}{m} \nabla \bm{F}^{(t)} \bm{1}} 
 =  -\frac{s\eta}{2} \norm{\nabla F(\bar{\x}^{t})}^2 \\
&~~~
+ \frac{s\eta}{2} \norm{\nabla F(\bar{\x}^{t}) - \frac{1}{m}\nabla \bm{F}^{(t)}\bm{1}}^2 
- \frac{s\eta}{2} \norm{\frac{1}{m}\nabla \bm{F}^{(t)}\bm{1}}^2\\
& \le  -\frac{s\eta}{2} \norm{\nabla F(\bar{\x}^{t})}^2 - \frac{s\eta}{2} \norm{\frac{1}{m}\nabla \bm{F}^{(t)}\bm{1}}^2 \\
&~~~+ \frac{s\eta L^2 }{2m}\sum_{i=1}^m \norm{\bar{\x}^{t} - \x_i^t}^2.   
\end{align*}

For term (B), we have
\begin{align*}
&\expect{\frac{\eta}{m} \iprod{\nabla F(\bar{\x}^{t})}{\sum_{i=1}^m s \nabla \ell_i^{(t,0)} - \sum_{k=0}^{s-1} \nabla \ell_i^{(t,k)}} \mid \calF^{t}} \\
& = \frac{\eta}{m} \sum_{i=1}^m\iprod{\nabla F(\bar{\x}^{t})}{ \expect{ s \nabla \ell_i^{(t,0)} - \sum_{k=0}^{s-1} \nabla \ell_i^{(t,k)}\mid \calF^{t}}} \\
& \overset{(\rma)}{\le} \frac{\eta^2 s^2}{2} \norm{\nabla F(\bar{x}^{t})}^2 \\
&~~~+ \underbrace{\frac{1}{2m s^2} \sum_{i=1}^m \expect{\norm{s \nabla \ell_i^{(t,0)} - \sum_{k=0}^{s-1} \nabla \ell_i^{(t,k)}}^2 \Big | ~\calF^{t}}}_{(\rmB.1)}, 
\end{align*}
where inequality $(\rma)$ holds because of Young's inequality.
From Lemma \ref{lemma: local step perturbation}, we bound term $(\rmB.1)$ as follows  
\begin{align*}
&\frac{1}{2m s^2} \sum_{i=1}^m \expect{\norm{s \nabla \ell_i^{(t,0)} - \sum_{k=0}^{s-1} \nabla \ell_i^{(t,k)}}^2 \mid \calF^{t}} \\
& \overset{(\rmb)}{\le} \frac{1}{2m s^2} \sum_{i=1}^m \expect{ \kappa^2 \eta^2 \binom{s}{2}^2 L^2\norm{\nabla \ell_{i}^{\pth{t,0}}}^2 \mid \calF^{t}} \\
& = \frac{\kappa^2 \eta^2 \binom{s}{2}^2 L^2}{2m s^2} \sum_{i=1}^m \expect{\norm{\nabla \ell_{i}^{\pth{t,0}} - \nabla F_i(\x_i^t) + \nabla F_i(\x_i^t)}^2 \mid \calF^{t}} \\
& \overset{(\rmc)}{\le} \kappa^2 \eta^2 L^2 \sigma^2\frac{s^2}{4} + \frac{\kappa^2 \eta^2 s^2 L^2}{4m} \sum_{i=1}^m\norm{\nabla F_i(\x_i^t)}^2\\
& \le \kappa^2 \eta^2 s^2 L^2 \frac{L^2}{m} \sum_{i=1}^m \norm{\x_i^t - \bar{\x}^t}^2 
+ \kappa^2 \eta^2 s^2 L^2 (\zeta^2 +\sigma^2)\\
&~~~+ \kappa^2 \eta^2 s^2 L^2\pth{\beta^2 + 1}\norm{\nabla F(\bar{\x}^t)}^2, 
\end{align*} 
where 
inequality $(\rmb)$ follows from Lemma~\ref{lemma: local step perturbation},
inequality $(\rmc)$ follows from Assumption \ref{ass: bounded variance client-wise}, 
and the last inequality holds because of Proposition \ref{proposition: average gradient to global gradient}. 
Combing the bounds of terms $(\rmA)$ and $(\rmB)$, we get 
\begin{align}
\label{eq: bound 1}
\nonumber
&\expect{\iprod{\nabla F(\bar{\x}^{t})}{- \frac{\eta}{m} \bm{G}^{(t)} \bm{1}} \mid \calF^{t}} \\
\nonumber
&\le- \qth{\frac{s\eta}{2} - \frac{\eta^2 s^2}{2} - \kappa^2 \eta^2 s^2 L^2\pth{\beta^2 + 1}} \norm{\nabla F(\bar{\x}^t)}^2 \\
\nonumber
&~~~ - \frac{s\eta}{2} \norm{\frac{1}{m}\nabla \bm{F}^{(t)}\bm{1}}^2 + \kappa^2 \eta^2 s^2 L^2 (\zeta^2 +\sigma^2)\\
&~~~ + \pth{\frac{s\eta L^2 }{2m} + \kappa^2 \eta^2 s^2 L^2 \frac{L^2}{m}} \sum_{i=1}^m \norm{\bar{\x}^{t} - \x_i^t}^2. 
\end{align}
\noindent{\em (b) Bounding $\expect{\norm{\frac{1}{m} \bm{G}^{(t)} \bm{1}}^2 \mid \calF^{t}}$.}
By adding and subtracting, we get
\begin{align*}
&\norm{\frac{1}{m} \bm{G}^{(t)} \bm{1}}^2  =  \norm{\frac{1}{m} \sum_{i=1}^m \sum_{k=0}^{s-1} \nabla \ell_i^{(t,k)}}^2 \\
& \le 2 \underbrace{\norm{\frac{1}{m} \sum_{i=1}^m \sum_{k=0}^{s-1} \pth{\nabla \ell_i^{(t,k)} - \nabla \ell_i^{(t,0)}}}^2}_{(\rmC)} + 2 \underbrace{\norm{\frac{s}{m} \sum_{i=1}^m   \nabla \ell_i^{(t,0)}}^2}_{(\rmD)}. 
\end{align*}    
\noindent For term $(\rmC)$, %
by Lemma \ref{lemma: local step perturbation}, we have
\begin{align*}
&\norm{\frac{1}{m} \sum_{i=1}^m \sum_{k=0}^{s-1} \pth{\nabla \ell_i^{(t,k)} - \nabla \ell_i^{(t,0)}}}^2 \\
&\le  \frac{\kappa^2 \eta^2 s^4 L^2}{4m}\sum_{i=1}^m \|\nabla \ell_i^{(t,0)}\|_2^2 \\
&\le \frac{\kappa^2 \eta^2 s^4 L^2}{2m} (\sum_{i=1}^m \norm{\nabla \ell_i^{(t,0)} - \nabla F_i(\x_i^t)}^2 + \sum_{i=1}^m \norm{\nabla F_i(\x_i^t)}^2) \\
& \overset{(\rmd)}{\le}  \frac{\kappa^2 \eta^2 s^4 L^2\sigma^2}{2} + \frac{\kappa^2 \eta^2 s^4 L^2}{2m} \sum_{i=1}^m \norm{\nabla F_i(\x_i^t)}^2,
\end{align*}
where inequality $(\rmd)$ holds because of Assumption \ref{ass: bounded variance client-wise}.
For term $(\rmD)$,  by Assumption \ref{ass: bounded variance client-wise}, we likewise have 
\begin{align*}
&\frac{s^2}{m^2} 
\expect{\|{\sum_{i=1}^m   \nabla \ell_i^{(t,0)}}\|_2^2 \Big| \calF^{t}}  
\le 
\frac{2 s^2}{m}
\pth{\sigma^2
+ 
\sum_{i=1}^m \norm{\nabla F_i(\x_i^t)}^2}.
\end{align*}
Combing the above upper bounds of (C) and (D)
and applying Proposition \ref{proposition: average gradient to global gradient}, we get 
\begin{align}
\nonumber
&\expect{\norm{\frac{1}{m} \bm{G}^{(t)} \bm{1}}^2 \mid \calF^{t}} 
\le 2 s^2\sigma^2\pth{\frac{2}{m} + \frac{\kappa^2 \eta^2 s^2 L^2}{2}} \\
\nonumber
&\qquad \quad \quad \quad+ 6 s^2L^2\pth{2 +\frac{\kappa^2 \eta^2 s^2 L^2}{2}} \frac{1}{m} \sum_{i=1}^m \norm{\x_i^t - \bar{\x}^t}^2 \\
\nonumber
&\qquad \quad \quad \quad +6 s^2\pth{\beta^2 + 1}\pth{2 +\frac{\kappa^2 \eta^2 s^2 L^2}{2}}\norm{\nabla F(\bar{\x}^t)}^2 \\
\label{eq: bound 2}
&\qquad \quad  \quad \quad + 6 s^2\zeta^2 \pth{2 +\frac{\kappa^2 \eta^2 s^2 L^2}{2}}. 
\end{align}
\noindent{\em (c) Putting them together.}
Combining \eqref{eq: bound 1} and \eqref{eq: bound 2}, we get
\begin{align*}
&\expect{F(\bar{\x}^{t+1})  -  F(\bar{\x}^{t}) \mid \calF^{t}}  
\le \kappa^2 \eta^2 s^2 L^2 (\zeta^2 +\sigma^2) \\
&~~~ - \frac{\eta s}{2} \norm{\frac{1}{m}\nabla \bm{F}^{(t)}\bm{1}}^2 + \frac{L\eta^2}{2}6 s^2\zeta^2 \pth{2 +\frac{\kappa^2  L^2}{2}} \\
&~~~ - \qth{\frac{\eta s}{2} - \frac{\eta^2 s^2}{2} - \kappa^2 \eta^2 s^2 L^2\pth{\beta^2 + 1}} \norm{\nabla F(\bar{\x}^t)}^2\\
&~~~ + \pth{\frac{s\eta L^2 }{2m} + \kappa^2 \eta^2 s^2 \frac{L^4}{m}} \sum_{i=1}^m \norm{\x_i^t - \bar{\x}^{t}}^2 \\
& ~~~ + \frac{L\eta^2}{2}6s^2L^2\pth{2 +\frac{\kappa^2 L^2}{2}} \frac{1}{m} \sum_{i=1}^m \norm{\x_i^t - \bar{\x}^t}^2 \\
& ~~~ +\frac{L\eta^2}{2}6 s^2\pth{\beta^2 + 1}\pth{2 +\frac{\kappa^2 L^2}{2}}\norm{\nabla F(\bar{\x}^t)}^2 \\
&~~~ + \frac{L\eta^2}{2} 2s^2\sigma^2\pth{\frac{2}{m} + \frac{\kappa^2 L^2}{2}}. 
\end{align*}

Assuming that $\eta\le {1}/[{108Ls(\beta^2+1)(1+\kappa^2L^2)}]$, the above displayed equation can be simplified as 
\begin{align*}
&\expect{F(\bar{\x}^{t+1})  -  F(\bar{\x}^{t}) \mid \calF^{t}}
\le -\frac{\eta s}{3} \norm{\nabla F(\bar{\x}^t)}^2 \\
&~~~+ \eta s\frac{L^2}{m} \sum_{i=1}^m \norm{\x_i^t - \bar{\x}^t}^2
+ \eta^2 s^2 6L\pth{\zeta^2+\sigma^2}\pth{1+L^2\kappa^2}. 
\end{align*} 
\end{proof}

\begin{proof}[\bf Proof of Lemma \ref{lemma: consensus}] 
Define $\Delta \bm{G}^{(r)} \triangleq \bm{G}^{(r)} -  \bm{G}_0^{(r)}$ and
$A_{r,t}\triangleq \prod_{\ell=r}^t W^{(\ell)} - \allones$.
The consensus error can be rewritten as 
\begin{align}
\nonumber
&\fnorm{\bm{X}^{(t)} \pth{\identity - \allones}}^2
= \fnorm{(\bm{X}^{(t-1)} - \eta \bm{G}^{(t-1)}) W^{(t-1)} \pth{\identity - \allones}}^2\\
\nonumber
&= \fnorm{- \eta \sum_{q=0}^{t-1} \bm{G}^{(q)} 
A_{q,t-1}
}^2 
\le 3\eta^2 \underbrace{\fnorm{
    \sum_{q=0}^{t-1} 
    \Delta \bm{G}^{(q)}
    A_{q,t-1}
    }^2}_{(\rmA)}\\
    \nonumber
    &+ 3\eta^2 
    \underbrace{\fnorm{\sum_{q=0}^{t-1}\pth{\bm{G}_0^{\pth{q}} - s\nabla \bm{F}^{\pth{q}}}
    A_{q,t-1}
    }}_{(\rmB)}\\
    \label{eq: conseneus iterative error}
    &+3\eta^2 s^2 \underbrace{\fnorm{\sum_{q=0}^{t-1}\nabla \bm{F}^{\pth{q}}
    A_{q,t-1}
    }^2}_{(\rmC)},
\end{align}
where the second equality follows from the fact that all clients are initiated at the same weights. 

\noindent{\em (a) Bounding $\expect{(\rmA)}$.}
The term $(\rmA)$ in Eq.\,\eqref{eq: conseneus iterative error} arises from multiple local steps. 
We have,
\begin{align}
\nonumber
&\expect{(\rmA) } 
\overset{(\rma)}{\le} \sum_{q=0}^{t-1} \rho^{t-q}\expect{
\fnorm{\Delta \bm{G}^{(q)}}^2 
} \\
\nonumber
&+ \sum_{q=0}^{t-1} \sum_{p=0, p\neq q}^{t-1}\expect{ 
\fnorm{
\Delta \bm{G}^{(p)}
A_{p,t-1}
}
\fnorm{
\Delta \bm{G}^{(q)}
A_{q,t-1}
} 
}\\
\nonumber
&\overset{(\rmb)}{\le}
\sum_{q=0}^{t-1} \rho^{t-q}\expect{
\fnorm{\Delta \bm{G}^{(q)}}^2 
} 
\\
\nonumber
&+ 
\sum_{q=0}^{t-1} \sum_{p=0, p\neq q}^{t-1} 
\frac{\sqrt{\rho}^{2t-p-q}}{2}
\expect{
{
\fnorm{
\Delta \bm{G}^{(p)}
}^2 
+
\fnorm{
\Delta \bm{G}^{(q)}
}^2 } }, 
\end{align}
where inequality $(\rma)$ follows from \eqref{eq: mathcha consensus},
inequality $(\rmb)$ holds because of Young's inequality.
Next, we bound the second term.
it follows that
\begin{align*}
    &\sum_{q=0}^{t-1} \sum_{p=0, p\neq q}^{t-1}  \frac{\sqrt{\rho}^{2t-p-q}}{2}\expect{
    {
    \fnorm{
    \Delta \bm{G}^{(p)}
    }^2 
    +
    \fnorm{
    \Delta \bm{G}^{(q)}
    }^2 } } \\ 
    & \le \sum_{q=0}^{t-1} \sum_{p=0}^{t-1} \frac{\sqrt{\rho}^{2t-p-q}}{2}
    \expect{
    {
    \fnorm{
    \Delta \bm{G}^{(p)}
    }^2 
    +
    \fnorm{
    \Delta \bm{G}^{(q)}
    }^2 } }
    \\
    &\le \frac{\sqrt{\rho}}{1-\sqrt{\rho}}\sum_{q=0}^{t-1}  \sqrt{\rho}^{t-q}\expect{
    \fnorm{
    \Delta \bm{G}^{(q)}
    }^2 }.
\end{align*}
In addition, since $\rho<1$, it holds that $\rho^{t-q} \le \sqrt{\rho}\rho^{\frac{t-q}{2}}$ for any $q\le t-1$. Thus, we have 
\begin{align}
\nonumber
\expect{(\rmA)}
& \le \sqrt{\rho}\sum_{q=0}^{t-1}\rho^{\frac{t-q}{2}} \expect{\fnorm{
\Delta \bm{G}^{(q)}
}^2 } \\
\nonumber
&\qquad \qquad + \frac{\sqrt{\rho}}{1-\sqrt{\rho}}\sum_{q=0}^{t-1}  \sqrt{\rho}^{t-q}\expect{\fnorm{
\Delta \bm{G}^{(q)}
}^2}\\
\label{eq: rho T1}
& \le \frac{2\sqrt{\rho}}{1-\sqrt{\rho}}\sum_{q=0}^{t-1}  \sqrt{\rho}^{t-q}\expect{\fnorm{\pth{\bm{G}^{\pth{q}} - \bm{G}_0^{\pth{q}}}}^2}. 
\end{align}
It remains to bound $\expect{\fnorm{\Delta \bm{G}^{\pth{q}}}^2 },$
\begin{align*}
&\expect{\fnorm{\Delta \bm{G}^{\pth{q}}}^2 } 
\overset{(\rmc)}{\le} \kappa^2 \eta^2 s^4 L^2 \expect{\fnorm{\bm{G}_0^{\pth{q}} - s\nabla \bm{F}^{\pth{q}} + s\nabla \bm{F}^{\pth{q}}  }^2 }\\
& \le 2 \kappa^2 \eta^2 s^4 L^2 \expect{\fnorm{\bm{G}_0^{\pth{q}} - s\nabla \bm{F}^{\pth{q}}}^2 } \\
&\qquad \qquad \qquad ~~\quad \quad + 2 \kappa^2 s^2 \eta^2 s^4 L^2 \expect{\fnorm{\nabla \bm{F}^{\pth{q}}}^2 }\\
& \le 2 \kappa^2 s^2 \eta^2 s^4 L^2 m \sigma^2 
+ 2 \kappa^2 s^2 \eta^2 s^4 L^2 \expect{\fnorm{\nabla \bm{F}^{\pth{q}}}^2 }, 
\end{align*}
where inequality ($\rmc$) follows from Lemma \ref{lemma: local step perturbation}, adding and subtracting.
Thus,
\begin{align*}
& \expect{(\rmA) } 
\le \frac{2\sqrt{\rho}}{1-\sqrt{\rho}}\sum_{q=0}^{t-1} 
\sqrt{\rho}^{t-q}\expect{\fnorm{\bm{G}^{\pth{q}} - \bm{G}_0^{\pth{q}}}^2 }\\
& \le \frac{4\kappa^2 s^2 \eta^2 s^4 L^2 m \sigma^2 \rho}{\pth{1-\sqrt{\rho}}^2} \\
&\qquad \qquad \quad + \frac{4\kappa^2 s^2 \eta^2 s^4 L^2\sqrt{\rho}}{1-\sqrt{\rho}}
\sum_{q=0}^{t-1} \sqrt{\rho}^{t-q}
\expect{\fnorm{\nabla \bm{F}^{\pth{q}}}^2}.
\end{align*}
\noindent{\em (b) Bounding $\expect{(\rmB)}$.}
\begin{align*}
\expect{(\rmB) } 
&\le \sum_{q=0}^{t-1}\rho^{t-q}\expect{
\fnorm{
\pth{\bm{G}_0^{\pth{q}} - s\nabla \bm{F}^{\pth{q}}}
}^2}
\le \frac{\rho m s^2 \sigma^2}{1-\rho} .
\end{align*}

\noindent{\em (c) Bounding $\expect{(\rmC)}$.}
Use a similar derivation as in \prettyref{eq: rho T1}, and we get
\begin{align*}
\expect{(\rmC)} 
&\le \frac{2\sqrt{\rho}}{1-\sqrt{\rho}}\sum_{q=0}^{t-1} \sqrt{\rho}^{t-q}\expect{\fnorm{\nabla \bm{F}^{\pth{q}}}^2 }.
\end{align*}
Furthermore, we have
\begin{align*}
&\sum_{t=0}^{T-1}\sum_{q=0}^{t-1} 
\sqrt{\rho}^{t-q}\expect{\fnorm{\nabla \bm{F}^{\pth{q}}}^2} 
=\sum_{t=0}^{T-2} \expect{\fnorm{\nabla \bm{F}^{\pth{t}}}^2} \sum_{q=1}^{T-1-t}\sqrt{\rho}^{q} \\
&\le \frac{\sqrt{\rho}}{\pth{1-\sqrt{\rho}}}\sum_{t=0}^{T-1}\expect{\fnorm{\nabla \bm{F}^{\pth{t}}}^2}.
\end{align*}

\noindent{\em (d) Putting them together.}
\begin{align}
\nonumber
&\frac{1}{mT}\sum_{t=0}^{T-1}\expect{\fnorm{\bm{X}^{\pth{t}} \pth{\identity - \allones}}^2}
\le 
3 \eta^2 s^2 \sigma^2 
\frac{
\rho
\pth{1 + \kappa^2 \eta^2 s^4 L^2}}{\pth{1 - \sqrt{\rho}}^2}\\
\nonumber
&~~~+ 
\pth{\frac{\kappa^2 \eta^2 s^4 L^2}{2} + 1}\frac{6 \eta^2 s^2 \rho}{mT\pth{1-\sqrt{\rho}}^2}\sum_{t=0}^{T-1}\expect{\fnorm{\nabla \bm{F}^{\pth{t}}}^2} \\
\nonumber
&\overset{(\rmd)}{\le}
\frac{9\rho}{(1-\sqrt{\rho})^2}\eta^2 s^2 \frac{1}{mT} \sum_{t=0}^{T-1} \fnorm{\nabla \bm{F}^{(t)}} 
+ \frac{6\rho \sigma^2}{(1-\sqrt{\rho})^2} \eta^2 s^2,
\end{align}
where we assume that $\eta \le \frac{1}{s^2 \kappa L}$ in inequality $(\rmd)$.
Choosing $\eta \le \frac{1-\sqrt{\rho}}{6Ls}$ and by Proposition \ref{proposition: average gradient to global gradient}, we have 
\begin{align*}
&\frac{1}{mT}\sum_{t=0}^{T-1}\expect{\fnorm{\bm{X}^{\pth{t}} \pth{\identity - \allones}}^2} 
 \le  
 \frac{12\rho \sigma^2}{(1-\sqrt{\rho})^2} \eta^2 s^2 \\
 &
 \frac{54(\beta^2+1)\rho \eta^2 s^2}{(1-\sqrt{\rho})^2} \frac{1}{mT} \sum_{t=0}^{T-1} \fnorm{\nabla \bm{F}(\bar{\bm{x}}^t)}^2 
+ \frac{54\rho \zeta^2}{(1-\sqrt{\rho})^2} \eta^2 s^2. 
\end{align*}
\end{proof}

\section{Conclusion}
In this paper,
we study federated learning
in the presence of
stochastic uplink communications
that are allowed to be simultaneously
{\em time-varying} and {\em unknown} to all parties 
in the distributed learning system.
We show that,
by using a simple quadratic counterexample in Proposition~\ref{proposition: nonuniform},
the seminal work FedAvg
is inherently biased from the global optimum
under non-\iid local data.
We propose FedPBC,
which leverages implicit gossiping by postponing the broadcast till the end of each global round,
is provable to reach a stationary point of the global non-convex objective, 
and converges at the optimal rate in the presence of smooth non-convex and stochastic objective gradients.
Extensive experiments have been provided over diversified unreliable patterns to corroborate our analysis.
Numerous directions are open for future research.
First, our work can be extended by studying unreliable bidirectional communication links.
We expect to incorporate different local optimization methods,
other than stochastic gradient descent,
and establish provable guarantees.
Finally,
it is also interesting to explore achieving the desired linear speedup property.
\newpage

\bibliography{citations/citation,citations/SM,citations/report}
\bibliographystyle{plainnat}

\newpage 

\appendix

\begin{center}   
\bf \LARGE
Appendix
\end{center}

\section{Proofs}
\label{app: proofs}
\begin{proof}[\bf Proof of Proposition \ref{proposition: nonuniform}]
At each client $i\in\calA^t,$ for each local step $k=0, \cdots, s-1$, we have
\begin{align*}
\x_i^{(t,k+1)} 
& = \pth{1-\eta}^{k+1} \x^t + \eta \bu_i \qth{\sum_{r=0}^k (1-\eta)^r}. 
\end{align*}
It follows that
\begin{align*}
&\x^{t+1}
= \indc{\calA_t=\emptyset} \x^t \\
&~~~+ \indc{\calA_t\not=\emptyset}\frac{1}{|\calA^t|}\sum_{i\in \calA^t}\pth{\pth{1-\eta}^{s} \x^t + \eta \bu_i \qth{\sum_{r=0}^{s-1} (1-\eta)^r}} \\
& = \x^t\indc{\calA_t=\emptyset} 
+ \pth{1 - \eta}^{s}\x^t\indc{\calA^t\neq \emptyset} \\%
&\qquad \qquad \qquad + \frac{\eta \sum_{i\in\calA^t}\bu_i \qth{\sum_{r=0}^{s-1} \pth{1-\eta}^r}\indc{\calA^t\neq\emptyset}}{\abth{\calA^t}}\\
&= \qth{\indc{\calA^t = \emptyset}+\pth{1-\eta}^{s} \indc{\calA^t \neq \emptyset}}\x^t \\
&\qquad \qquad \qquad + \qth{1-(1-\eta)^s}\frac{\indc{\calA^t \neq \emptyset}}{\abth{\calA^t}}\sum_{i\in \calA^t}\bu_i. 
\end{align*}    
Taking expectation with respect to $\calA^t$, we get 
\begin{align*}
&\expect{\x^{t+1} \mid \calA^t}  = \qth{\prob{\calA^t = \emptyset}+\pth{1-\eta}^{s} \prob{\calA^t \neq \emptyset}} \x^t \\ %
&~~~+ \qth{1-(1-\eta)^s} \expect{\frac{\sum_{i\in \calA^t}\bu_i}{\abth{\calA^t}}\Big| \calA^t \neq \emptyset} \prob{\calA^t \neq \emptyset}\\
& = \pth{\prod_{i=1}^m\pth{1-p_i} + \qth{1 -\prod_{i=1}^m\pth{1-p_i}}\pth{1-\eta}^s}\x^t \\
&+ \qth{1-(1-\eta)^s} [{1-\prod_{i=1}^m\pth{1-p_i}}]\expect{\frac{\sum_{i\in \calA^t}\bu_i}{\abth{\calA^t}}\Big| \calA^t \neq \emptyset}. 
\end{align*}
Following from the fact that $p_i^t=p_i$ for all $t$ at all clients, 
$\mathbb{E}[{\frac{1}{\abth{\calA^t}}\sum_{i\in \calA^t}\bu_i | \calA^t \neq \emptyset}] = \mathbb{E}[{\frac{1}{\abth{\calA^1}}\sum_{i\in \calA^1}\bu_i | \calA^1 \neq \emptyset}]$ for all $t$. 
Unrolling the above displayed equation until time 0 and applying the full expectation up to time $t+1$, we have 
\begin{align}
\expect{\x^{t+1}} 
& = \pth{1 - {\rma}^{t+1}} \expect{\expect{\frac{1}{\abth{\calA^1}}\sum_{i\in \calA^1}\bu_i \Big| \calA^1 \neq \emptyset}},
\label{eq: At full expectation}
\end{align}
where  $\x^0=\bm{0}$, and 
\begin{align*}
\rma& \triangleq \prod_{i=1}^m\pth{1-p_i} + \qth{1 -\prod_{i=1}^m\pth{1-p_i}}\pth{1-\eta}^s. 
\end{align*}
Notably, $\rma<1$,
it holds that
$
\lim_{t\rightarrow\infty} (1 - \rma^{t+1}) = 1. 
$
Let $X_i = \indc{i\in \calA^1}$ for each $i\in [m]$. We have 
\begin{align*}
\expect{\frac{\sum_{i\in \calA^1}\bu_i}{\abth{\calA^1}} \Big| \calA^1 \neq \emptyset} 
& =\expect{\frac{\sum_{i=1}^m X_i \bu_i}{\sum_{j=1}^m X_j} \Big| \sum_{j=1}^m X_j \neq 0} \\
&= \sum_{i=1}^m \bu_i\expect{ \frac{X_i}{\sum_{j=1}^m X_j} \Big|  \sum_{j=1}^m X_j \neq 0}.
\end{align*}
By the law of total expectation and the convention that $\frac{0}{0}=0$, we know that
\begin{small}
\begin{align*}
\expect{\frac{X_i}{\sum_{j=1}^m X_j}}
& = 
\expect{\frac{X_i}{\sum_{j=1}^m X_j} \Big| \sum_{j=1}^m X_j \neq 0}
\prob{\sum_{j=1}^m X_j \neq 0}
+
0 \\
&=
\expect{\frac{X_i}{\sum_{j=1}^m X_j} \Big| \sum_{j=1}^m X_j \neq 0}
\prob{\sum_{j=1}^m X_j \neq 0}. 
\end{align*}
\end{small}
Hence,%
\begin{align*}
& \expect{\frac{X_i}{\sum_{j=1}^M X_j}  \Big| \sum_{j=1}^M X_j \neq 0}
=
\frac{\expect{\frac{X_i}{\sum_{j=1}^m X_j}}}{1 - \prod_{i=1}^m (1-p_i)}.
\end{align*}
Additionally, 
\begin{align}
\nonumber
\expect{ \frac{X_i}{\sum_{i=1}^m X_i}} 
&= \prob{X_i = 1}\expect{ \frac{X_i}{\sum_{j=1}^m X_j}\Big| X_i = 1} 
+ 0\\
&= p_i\expect{ \frac{1}{1 + \sum_{j\in [m]\setminus \sth{i}} X_j}\Big| X_i = 1}.
\label{eq: Xi full expectation}
\end{align}
Next, we show that
\begin{align}
\nonumber 
&\expect{ \frac{1}{1 + \sum_{j\in [m]\setminus \sth{i}} X_j}\Big| X_i = 1} \\
&\qquad \qquad \qquad \qquad =
1+\sum_{j=2}^{m} \pth{-1}^{j+1} \frac{1}{j}\sum_{S\in \calB^i_j} \prod_{z\in S} p_z,
\label{eq: induction claim in counterexample}
\end{align}
where $\calB^i_j \triangleq \sth{S \Big|S\subseteq [m]\setminus\sth{i}, \abth{S} = j-1}$.
Without loss of generality, assume $i = m$.
Define $\bar{S} \triangleq [m] \setminus S$
\begin{align}
\nonumber
&\expect{ \frac{1}{1 + \sum_{j\in [m]\setminus \sth{m}} X_j}\Big| X_m = 1}
=
\expect{ \frac{1}{1 + \sum_{j\in [m-1]} X_j}} \\
\nonumber
&\triangleq
\sum_{j=1}^{m}
\frac{1}{j} \prob{\abth{\calA^1 \setminus \sth{m}} = j-1} \\
&=
\sum_{j=1}^{m}
\frac{1}{j}
\sum_{S\in \calB_j} \prod_{x \in \bar{S}}(1-p_x) \prod_{z\in S} p_z.
\label{eq: counterexample expression}
\end{align}
Then, we show that~\prettyref{eq: induction claim in counterexample} and~\prettyref{eq: counterexample expression} are equivalent.
The degree coefficient of polynomial 0 (\ie, when $\abth{S}=0$)
relates only to $j\in \sth{1}$: $\prod_{k=1}^{m-1} (1 - p_k)$,
where we select all the ones in parentheses.
Thus, the coefficient of the terms in the degree of polynomial 0 is $1$.
The degree coefficient of polynomial 1 (\ie, when $\abth{S}=1$).
corresponds to $j\in \sth{1,2}$:
\begin{align}
\label{eq: j=1 in power 1}
&\prod_{k=1}^{m-1} (1 - p_k)~(j=1); \\
\label{eq: j=2 in power 1}
\frac{1}{2}\sum_{k=1}^{m-1} p_k &\prod_{x\in [m-1]\setminus \sth{k}} (1 - p_x)~(j=2).
\end{align}
Take the coefficient of $p_1$ as an example.
In \eqref{eq: j=1 in power 1},
to get $p_1$,
we select $p_1$ from $(1-p_1)$ and all the ones from the rest parentheses,
which yields $-1\binom{1}{0}$.
In addition,
in \eqref{eq: j=2 in power 1},
the coefficient is $\frac{1}{2}{1 \choose 1}$.
They add up to $-1 + \frac{1}{2} = -\frac{1}{2}$.
For a general degree coefficient of polynomial $K$ (\ie, when $\abth{S}=K$), 
by using a similar argument,
the coefficient is
$(-1)^{K} \qth{\sum_{y=0}^{K} \frac{(-1)^{y}}{y+1}{K \choose y}}$,
which can be simplified as
\begin{align*}
&\pth{-1}^K
\sum_{y=0}^{K} \frac{(-1)^{y}}{y+1}{K \choose y}
=
\pth{-1}^{K}
\sum_{y=0}^{K} \frac{(-1)^{y}}{y+1}\frac{K!}{y!(K-y)!} \\
&=
\pth{-1}^K
\frac{1}{K+1}
\sum_{y=0}^{K} (-1)^{y}\frac{(K+1)!}{(y+1)!(K-y)!} \\
&=
\frac{\pth{-1}^{K+1}}{K+1}
\sum_{y=0}^{K} (-1)^{y+1}{K+1 \choose y+1} \\
&=
\frac{\pth{-1}^{K+1}}{K+1}
\qth{(-1+1)^{K+1} - (-1)^0}
=
\frac{\pth{-1}^K}{K+1}.
\end{align*}
Combining the above yields \eqref{eq: induction claim in counterexample}.
Finally, we plug Eq.~\prettyref{eq: Xi full expectation} in Eq.~\prettyref{eq: At full expectation} and get
\begin{align*}
&\lim_{t\diverge} \expect{\x^{t+1}}
=
\lim_{t\diverge}
\expect{
\sum_{i=1}^m
\bu_i
\expect{\frac{X_i}{\sum_{j=1}^m X_j} \Big| \calA^1 \neq \emptyset}
} \\
&=
\sum_{i=1}^m
\frac{\bu_i p_i \pth{1+\sum_{j=2}^{m} \pth{-1}^{j+1} \frac{1}{j}\sum_{S\in \calB_j} \prod_{z\in S} p_z}}{1 - \prod_{i=1}^m (1-p_i)},
\end{align*}
where $\calB_j \triangleq \sth{S \Big|S\subseteq [m]\setminus\sth{i}, \abth{S} = j-1}$. 
\end{proof}

\begin{proof}[\bf Proof of Lemma \ref{lemma: ergodicity}]
For ease of exposition, in this proof we drop time index $t$. 
We first get the explicit expression for $\expect{W^2_{jj^{\prime}}\mid \calA \neq \emptyset}$. 
Suppose that $\calA\not=\emptyset$. 
We have 
\begin{align*}
W^2_{jj^{\prime}} & = \sum_{k=1}^m W_{jk}W_{j^{\prime}k} \\
&= W_{jj}W_{j^{\prime}j} + W_{jj^{\prime}}W_{j^{\prime}j^{\prime}} + \sum_{k\in [m]\setminus \{j, j^{\prime}\}} W_{jk}W_{j^{\prime}k}.   
\end{align*}
When $k\not=j$ and $k\not=j^{\prime}$ by Eq.\,\eqref{eq: gossiping matrix}, we have 
\begin{align*}
W_{jk}W_{j^{\prime}k} & = \frac{1}{|\calA|^2} \indc{j\in \calA} \indc{j^{\prime}\in \calA}\indc{k\in \calA}.     
\end{align*}
In addition, we have $W_{jj}W_{j^{\prime}j}  = \frac{1}{|\calA|^2} \indc{j\in \calA}\indc{j^{\prime}\in \calA},$ and $W_{j^{\prime}j^{\prime}}W_{jj^{\prime}} = \frac{1}{|\calA|^2} \indc{j\in \calA}\indc{j^{\prime}\in \calA}$.  
Thus, 
\begin{itemize}
\item 
For  $j \neq j^\prime$, we have
\begin{align*}
& W^2_{jj^{\prime}} = \sum_{k=1}^m W_{jk}W_{j^{\prime}k} 
= \frac{1}{|\calA|}\indc{j\in \calA} \indc{j^{\prime}\in \calA};
\end{align*}
\item
For $j = j^\prime$, we have
\begin{align*}
& W^2_{jj} = \frac{1}{|\calA|}\indc{j\in \calA} + \pth{1-\indc{j\in \calA}}.
\end{align*}
\end{itemize}
\noindent{\em (a) The general case where $p_i^t \ge c$.}
In the special case where $\calA = \emptyset$, we simply have $W = \identity$ by the algorithmic clauses.
Therefore,
$\expect{W_{j j^\prime} \mid \calA = \emptyset} \ge 0$ holds for any pair of $j,j^\prime  \in [m]$.
It follows, by the law of total expectation and for all $j,j^\prime \in [m]$, that
\begin{align}
\nonumber
\expect{W_{j j^\prime}}
&=
\expect{W_{j j^\prime} \mid \calA = \emptyset}
\prob{\calA = \emptyset} \\
\nonumber
&\qquad \qquad \qquad +
\expect{W_{j j^\prime} \mid \calA \neq \emptyset}
\prob{\calA \neq \emptyset} \\
&\ge
\expect{W_{j j^\prime} \mid \calA \neq \emptyset}
\prob{\calA \neq \emptyset}.
\label{eq: W matrix total expectation}
\end{align}
\begin{itemize}[leftmargin=*]
\item 
For $j \neq j^\prime$, it holds that
\begin{align*}
&\expect{W^2_{jj^{\prime}}\mid \calA \neq \emptyset} 
=
\expect{\frac{1}{|\calA|}\indc{j\in \calA} \indc{j^{\prime}\in \calA} \Big| \calA \neq \emptyset} \\
&\qquad \qquad \overset{(\rma)}{\ge} 
\expect{\frac{1}{m}\indc{j\in \calA} \indc{j^{\prime}\in \calA} \Big| \calA \neq \emptyset}
= 
\frac{p_j p_{j^{\prime}}}{m} 
\ge \frac{c^2}{m},
\end{align*}
where $(\rma)$ holds because $\abth{\calA} \le m$
;
\item
For $j = j^\prime$, it holds that
\begin{align*}
&\expect{W^2_{jj}\mid \calA\neq\emptyset} 
= 
\expect{\frac{1}{|\calA|}\indc{j\in \calA} + \pth{1-\indc{j\in \calA}} \Big| \calA \neq \emptyset} \\
&\ge
\expect{\frac{1}{m}\qth{\indc{j\in \calA} + \pth{1-\indc{j\in \calA}}} \Big| \calA \neq \emptyset} 
=
\frac{1}{m}
\ge
\frac{c^2}{m}.
\end{align*}
\end{itemize}
Recall that $M = \expect{W}$.
Next, we show that each element of $M$ is lower bounded.
\begin{align*}
M_{jj^{\prime}}
\ge 
\expect{W_{j j^\prime}^2\mid \calA \neq \emptyset}
\prob{\calA \neq \emptyset}
\ge \frac{c^2}{m} \qth{1-\pth{1-c}^m}.
\end{align*} 
We note that $\rho (t) = \lambda_2 (M)$,
where $\lambda_2$ is the second largest eigenvalue of %
matrix $M$. 
A Markov chain with $M$ as the transition matrix is ergodic as the chain is (1) {\it irreducible}: $M_{j j^\prime}\ge  \frac{c^2}{m}\qth{1-\pth{1-c}^m}>0$ for $j,j^\prime \in[m]$ and (2) {\it aperiodic} (it has self-loops).  
In addition, $W$ matrix is by definition doubly-stochastic. 
Hence, $M$ has a uniform stationary distribution
$
\pi = \frac{1}{m}\Indc^\top.
$
Furthermore, the irreducible Markov chain is reversible since 
it holds
for all the states that
$
\pi_i M_{ij} = \pi_j M_{ji}.
$
The conductance of a reversible Markov chain \cite{jerrum1988conductance} with  a transition matrix $M$ can be bounded by
\begin{align*}
&\Phi(M) 
= \min_{\sum_{i\in\calS} \pi_i \le \frac{1}{2}} \frac{\pi_i \sum_{i\in\calS, j\notin \calS} M_{ij}}{\sum_{i\in \calS} \pi_i}\\
&\quad \ge \frac{\pth{\frac{c}{m}}^2\qth{1-\pth{1-c}^m}\abth{\calS}\abth{\bar{\calS}}}{\frac{\abth{\calS}}{m}} = \frac{c^2\qth{1-\pth{1-c}^m}}{m} \abth{\bar{\calS}},
\end{align*}
where
$
\abth{\bar{\calS}} = m - \abth{\calS} \ge \frac{m}{2}.
$
From Cheeger's inequality, we know that
$
\frac{1 - \lambda_2}{2} \le \Phi(M) \le \sqrt{2 \pth{1-\lambda_2}}.
$
Finally, we have
\begin{align*}
\Phi(M) 
&\ge \frac{c^2\qth{1-\pth{1-c}^m}}{m} \abth{\bar{\calS}} \ge \frac{c^2\qth{1-\pth{1-c}^m}}{2}.
\end{align*}
Thus,
$
\rho(t) = \lambda_2 \le 1 - \frac{\Phi^2\pth{M}}{2} \le 1 - \frac{c^4\qth{1-\pth{1-c}^m}^2}{8}.
$

\vskip 1.5\baselineskip
\noindent{\em (b) Select $k$ clients uniformly at random.}
In each round,
the server picks $k$ clients uniformly at random.
Consequently,
different from the general case where $\abth{\calA}$ is a random variable,
it holds that $\abth{\calA} = k$ and $\calA\neq\emptyset$.
In addition,
$c \triangleq \frac{k}{m}$.
After a similar argument as in the first case,
it holds that $M_{j j^\prime} \ge \frac{c^2}{k}$.
The conductance of the reversible Markov chain with a transition matrix $M$ can be bounded by
$\Phi(M) \ge \frac{c^2}{k} \abth{\bar{\calS}} \ge \frac{c}{2}$.
Finally, we have
$
\rho(t) = \lambda_2 \le 1 - \frac{\Phi^2\pth{M}}{2} \le 1 - \frac{c^2}{8}.
$
\end{proof}
\begin{proof}[\bf Proof of~\prettyref{prop: staleness term}]
In our work,
the probabilities $p_i^t \ge c$.
Therefore,
define $Y_{\min}$ as the random variable of the ordinary geometric distribution with success probability $c$.
We have $\expect{Y_{\min}} = 1 / c$.
\cite[Theorem 3.2]{mandelbaum2007nonhomogeneous} tells us that $\expect{t - \tau_i(t)} \le \expect{Y_{\min}} = 1 /c $.
\end{proof}
\begin{proof}[\bf Proof of Theorem \ref{thm: main}]
In this proof,
we combine all the above intermediate results to show the final theorem.

\noindent{\em (a) Taking expectation over the remaining randomness and a telescoping sum.}
\begin{align*}
&\frac{1}{T}\sum_{t=0}^{T-1}\expect{F(\bar{\x}^{t+1})  -  F(\bar{\x}^{t}) }
\le 
- \frac{s\eta}{3} 
\frac{1}{T}\sum_{t=0}^{T-1}
\expect{\norm{\nabla F(\bar{\x}^t)}^2} \\
&
+ 6 L \eta^2 s^2 \pth{\kappa^2 L^2 + 1}\pth{\sigma^2 + \zeta^2} 
+ \eta s \frac{L^2}{mT}\sum_{t=0}^{T-1}\expect{\norm{\x_i^t - \bar{\x}^t}^2},
\end{align*}
where inequality $(a)$ holds because of~\prettyref{ass: lower bounds}.

\noindent{\em (b) Plugging in Lemma~\ref{lemma: consensus} and~\prettyref{ass: lower bounds}.}
\begin{align}
\nonumber
&\frac{F^\star  -  \expect{F(\bar{\x}^{0})}}{T}\\
\nonumber
& \le 
9 \eta^2 s^2 L
\qth{
\kappa^2 L^2 + 1
+ 
16 \eta s^2
\frac{\rho s L}{\pth{1 - \sqrt{\rho}}^2}
}
\pth{\sigma^2 + \zeta^2}
\\
\label{eq: main theorem intermediate}
&- \frac{s\eta}{3} 
\pth{1 - 162 \eta^2 s^2 \frac{\rho \pth{\beta^2 + 1} L^4}{\pth{1-\sqrt{\rho}}^2}}
\frac{1}{T}\sum_{t=0}^{T-1}
\expect{\norm{\nabla F(\bar{\x}^t)}^2} 
.
\end{align}
We know from 
$\eta 
\le
\frac{1-\sqrt{\rho}}{108 L^2 s^3 \pth{\beta^2+1} \pth{1+\kappa^2 L^2}}
\le
\frac{1 - \sqrt{\rho}}{18 (\beta^2 + 1) L^2 s}$ that
\begin{align*}
&1 - 162 \eta^2 s^2 \frac{\rho \pth{\beta^2 + 1} L^4}{\pth{1-\sqrt{\rho}}^2} \\
&\qquad \qquad \ge 
1 - \frac{162 \rho \pth{\beta^2 + 1} L^4}{\pth{1-\sqrt{\rho}}^2} \frac{\pth{1-\sqrt{\rho}}^2}{324 \pth{\beta^2 + 1}^2 L^4}\ge
\frac{1}{2}.
\end{align*}
In addition, we also have
$
\kappa^2 L^2 + 1
+ 
16 \eta s^3
\frac{\rho L}{\pth{1 - \sqrt{\rho}}^2}
\le
\kappa^2 L^2 + 1
+ 
\frac{1}{1 - \sqrt{\rho}}.
$
Therefore, 
rearrange the terms in \eqref{eq: main theorem intermediate}, 
it follows that
\begin{align*}
&\frac{1}{T}\sum_{t=0}^{T-1}
\expect{\norm{\nabla F(\bar{\x}^t)}^2}
\le
\frac{6 \pth{F(\bar{\x}^{0}) - F^\star}}{\eta s T} \\
&\qquad \qquad \qquad+
54 \eta s L
\pth{
\kappa^2 L^2 + 1
+
\frac{1}{1 - \sqrt{\rho}}}
\pth{\sigma^2 + \zeta^2}
.
\end{align*}
\end{proof}

\section{Experimental Setup}
\label{app: exp}
\noindent{\bf Hardware and Software Setups.}
The simulations are performed on a private cluster with 64 CPUs, 500 GB RAM and 8 NVIDIA A5000 GPU cards.
We code the experiments based on PyTorch 1.13.1 \cite{paszke2019pytorch} and Python 3.7.16.

\vskip.5\baselineskip
\noindent{\bf Neural Network and Hyper-parameter Specifications.}
\label{app: hyperparameter}
We initialize the customized CNNs using the Kaiming initialization.
A decaying learning rate schedule $\eta = \eta_0/ {\sqrt{({t}/{10}) + 1}}$ is adopted.
The initial local learning rate $\eta_0$ and the global learning rate $\eta_g$ 
are searched, 
based on the best performance after $500$ global rounds, 
over two grids $\sth{0.1, 0.05, 0.01, 0.005, 0.001, 0.0005}$ and $\sth{0.5,1,1.5,5,10,50}$, respectively.
We set $\beta = 0.01$ for F3AST \cite{ribero2022federated}., 
which is tuned over a grid of 
\[
    \sth{1, 0.5, 0.1, 0.05, 0.01, 0.005} \times 10^{-2}.
\]

\vskip.5\baselineskip
\noindent{\bf Missing algorithm descriptions.}
In this section, 
we specify the missing essential hyperparameters for specific algorithm implementations.
As recommended by \cite{wang2023lightweight}, 
we choose $K=50$ for FedAU without further specification.
Note that $K$ is an algorithmic hyperparameter in FedAU. 
Adopting the setup in \cite{ribero2022federated},
we set the communication constraint to be $10$ clients for F3AST.

\vskip.5\baselineskip
\noindent{\bf Datasets.}
All the datasets we evaluate contain 10 classes of images.
Some data enhancement tricks that are standard in training image classifiers are applied during training.
Specifically,
we apply random cropping to all datasets.
Furthermore, random horizontal flipping is applied to CIFAR-10 and CINIC-10.
{SVHN \cite{netzer2011reading}}
dataset contains 32$\times$32 colored images of 10 different number digits.
In total,
there are 73257 train images and 26032 test images.
{CIFAR-10 \cite{krizhevsky2009learning}}
dataset contains 32$\times$32 colored images of 10 different objects.
In total,
there are 50000 train images and 10000 test images.
{CINIC-10\cite{darlow2018cinic}} 
dataset contains 32$\times$32 colored images of 10 different objects.
In total,
there are 90000 train images and 90000 test images.

\vskip.5\baselineskip
\noindent{\bf Constructions of Markov transition probabilities.}
Recall that the link status in Markovian unreliable scheme is dictated by a Markov chain, whose initial states are based on $\mathsf{Bernoulli}(p_i^t)$.
\prettyref{fig: markovian transition probability} plots the Markov chain.
Let $q_i^t$ and $q_i^{t\star}$ define the transition probability from the ``ON'' state to the ``OFF'' state and from the ``OFF'' state to the ``ON'' state, respectively.
In the experiments, 
we aim to construct $q_i^t$ and $q_i^{t\star}$ so that a stationary distribution is met as
\begin{align}
\label{eq: markov equilibrium}
q_i^t \cdot p_i^t & = q_i^{t\star} \cdot \pth{1 - p_i^t}.
\end{align}
Concretely,
we first assume that $q_i^{t\star} = 0.05$ is an external choice.
If $q_i^{t\star} \cdot \pth{1 - p_i^t} > {p_i^t}$,
we adjust $q_i^t$ and $q_i^{t\star}$ to ensure \eqref{eq: markov equilibrium}.
Please find the details in~\prettyref{tab: markov transition probs}.

\begin{figure}[!t]
\begin{minipage}[b]{.5\textwidth}
\centering
\begin{tabular}{c|cc}
\toprule
Transition  probabilities
 & \multirow{2}{*}{$q_i^{t\star}$} & \multirow{2}{*}{$q_i^t$}\\
 \cline{1-1}
 Conditions & & \\
 \midrule
$q_i^{t\star} \cdot \pth{1 - p_i^t} \le {p_i^t}$ & 0.05 &  $0.05 \cdot \frac{1-p_i^t}{p_i^t}$\\
$q_i^{t\star} \cdot \pth{1 - p_i^t} > {p_i^t}$ & $\frac{p_i^t}{1-p_i^t}$ & 1\\
\bottomrule
\end{tabular}
\captionsetup{type=table}
\caption{\small The construction of $q_i^t$ and $q_i^{t\star}$.}
\label{tab: markov transition probs}
\end{minipage}
\begin{minipage}[b]{.5\textwidth}
\centering
\resizebox{\linewidth}{!}{
\tikzset{every picture/.style={line width=0.75pt,font=\LARGE}} %
\begin{tikzpicture}[x=0.75pt,y=0.75pt,yscale=-1,xscale=1]

\draw   (150,144.5) .. controls (150,130.69) and (161.19,119.5) .. (175,119.5) .. controls (188.81,119.5) and (200,130.69) .. (200,144.5) .. controls (200,158.31) and (188.81,169.5) .. (175,169.5) .. controls (161.19,169.5) and (150,158.31) .. (150,144.5) -- cycle ;
\draw   (315,144.5) .. controls (315,130.69) and (326.19,119.5) .. (340,119.5) .. controls (353.81,119.5) and (365,130.69) .. (365,144.5) .. controls (365,158.31) and (353.81,169.5) .. (340,169.5) .. controls (326.19,169.5) and (315,158.31) .. (315,144.5) -- cycle ;
\draw    (189.63,123.12) .. controls (228.63,93.87) and (295.8,103.7) .. (317.55,128.67) ;
\draw [shift={(319.13,130.62)}, rotate = 233.02] [fill={rgb, 255:red, 0; green, 0; blue, 0 }  ][line width=0.08]  [draw opacity=0] (7.14,-3.43) -- (0,0) -- (7.14,3.43) -- (4.74,0) -- cycle    ;
\draw    (193.83,166.45) .. controls (235.48,189.62) and (281.12,189.39) .. (319.63,161.12) ;
\draw [shift={(190.63,164.62)}, rotate = 30.47] [fill={rgb, 255:red, 0; green, 0; blue, 0 }  ][line width=0.08]  [draw opacity=0] (7.14,-3.43) -- (0,0) -- (7.14,3.43) -- (4.74,0) -- cycle    ;
\draw    (152.63,130.62) .. controls (128.15,134.94) and (128.1,150.32) .. (149.36,155.08) ;
\draw [shift={(152.13,155.62)}, rotate = 189.46] [fill={rgb, 255:red, 0; green, 0; blue, 0 }  ][line width=0.08]  [draw opacity=0] (7.14,-3.43) -- (0,0) -- (7.14,3.43) -- (4.74,0) -- cycle    ;
\draw    (362.63,131.12) .. controls (381.55,130.15) and (392.47,150.82) .. (364.81,155.71) ;
\draw [shift={(362.13,156.12)}, rotate = 352.65] [fill={rgb, 255:red, 0; green, 0; blue, 0 }  ][line width=0.08]  [draw opacity=0] (7.14,-3.43) -- (0,0) -- (7.14,3.43) -- (4.74,0) -- cycle    ;

\draw (245,75.4) node [anchor=north west][inner sep=0.75pt]    {$q_{i}^{t\star}$};
\draw (244,150.4) node [anchor=north west][inner sep=0.75pt]    {$q_{i}^{t}$};

\draw (155,135) node [anchor=north west][inner sep=0.75pt]   [font=\normalsize][align=center] {{\Large OFF} };
\draw (325.5,135) node [anchor=north west][inner sep=0.75pt]   [font=\normalsize][align=center] {{\Large ON}};
\draw (65.5,130.5) node [anchor=north west][inner sep=0.75pt]    {$1-q_{i}^{t\star}$};
\draw (386.5,130.5) node [anchor=north west][inner sep=0.75pt]    {$1-q_{i}^{t}$};
\end{tikzpicture}}
\caption{\small An illustration of the Markovian transition probabilities.}
\label{fig: markovian transition probability}
\end{minipage}
\end{figure}

\begin{figure*}[!t]
    \centering
    \begin{subfigure}[b]{\textwidth}
    \centering
    \includegraphics[width=.85\linewidth]{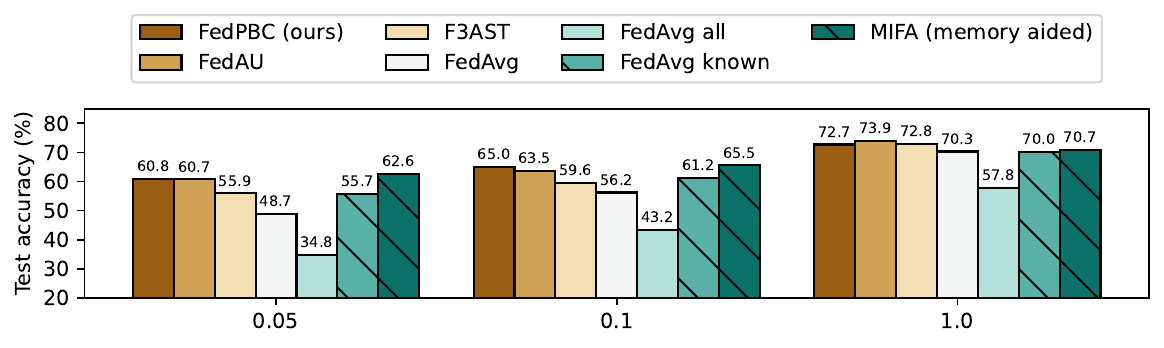}
    \vskip -.5\baselineskip
    \caption{The impact of data heterogeneity $\alpha$.}
    \label{fig: impact of alpha}
    \end{subfigure}
    \begin{subfigure}[b]{\textwidth}
    \centering
    \includegraphics[width=.85\linewidth]{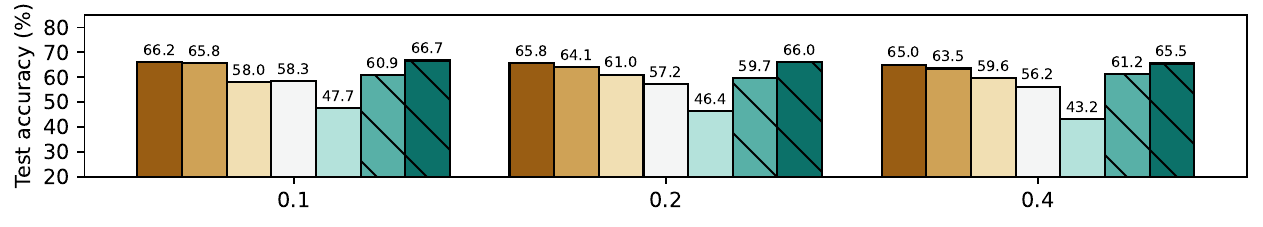}
    \vskip -.5\baselineskip
    \caption{The impact of probability fluctuation $\gamma$.}
    \label{fig: impact of gamma}
    \end{subfigure}
        \begin{subfigure}[b]{\textwidth}
        \centering
    \includegraphics[width=.85\linewidth]{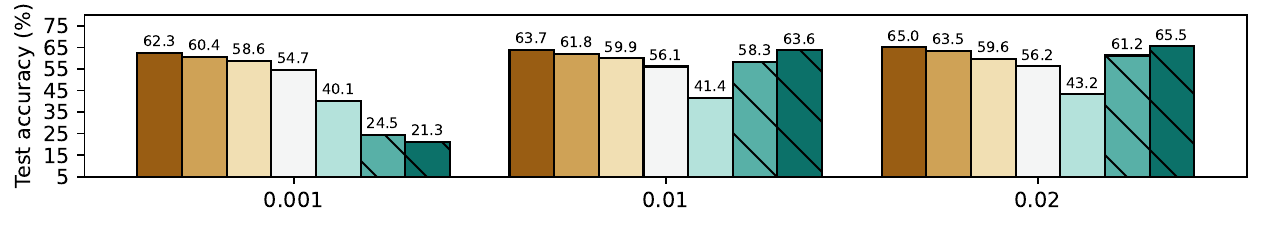}
    \vskip -.5\baselineskip
    \caption{The impact of a cutting-off lower bound $\delta$. 
    }
    \label{fig: impact of delta}
    \end{subfigure}
        \begin{subfigure}[b]{\textwidth}
        \centering
    \includegraphics[width=.85\linewidth]{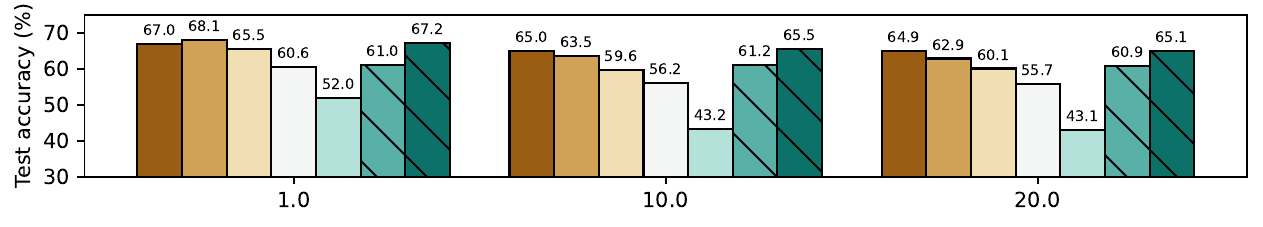}
    \vskip -.5\baselineskip
    \caption{The impact of contribution heterogeneity $\sigma_0$.}
    \label{fig: impact of sigma}
    \end{subfigure}
    \caption{
    \small
    The test accuracies in the ablation experiments.
    In each plot,
    only one system design parameter is changed.
    The others remain the same as in Table~\ref{tbl: exp main text}.
    All experiments are evaluated on the CIFAR-10 dataset under Bernoulli with {\em time-varying} unreliable uplinks.
    The bars with backslashes refer to the algorithms requiring extra memory or {\em known} historical statistics.
    }
    \label{fig: ablation study}
\end{figure*}
\noindent{\bf Ablation Experiments.}
In this part, 
we conduct ablation experiments to study the impact of different parameters on the performance of FedPBC and the other baseline algorithms.
Specifically, we evaluate all algorithms on the CIFAR-10 dataset under the Bernoulli unreliable communication scheme with {\em time-varying} $p_i^t$'s.
In any set of experiments, only one system design parameter is changed, 
while the others remain the same as in Table~\ref{tbl: exp main text}.
We report the mean test accuracy over the last 100 rounds in bar plots in~\prettyref{fig: ablation study}.
Algorithms are divided into two groups:
those with additional memory or {\em known} historical statistics (bars with backslashes) and those without.
It is observed that FedPBC outperforms the baseline algorithms {\em not} aided by memory in almost all cases (except when $\alpha = 1.0$ by FedAU in \prettyref{fig: impact of alpha} and $\sigma_0 = 1.0$ by FedAU in \prettyref{fig: impact of sigma}.)
The reason why FedPBC trails behind FedAU in the above two cases is worth further investigation.
Compared to memory-aided algorithms,
although MIFA occasionally dwarfs FedPBC,
the benefit margin is lower than 2\% in test accuracy.

\noindent{\bf Impact of data heterogeneity $\alpha$.}
In the presence of more homogenous local data,
\ie,
a larger $\alpha$,
the bias phenomenon gradually disappears as the local objectives become interchangeable,
which is confirmed by \prettyref{fig: impact of alpha} from the on-par performance of almost all algorithms when $\alpha = 1.0$.

\noindent{\bf Impact of fluctuation $\gamma$.}
The magnitude of the sine function is defined as $\gamma$ and thus governs the fluctuations of $p_i^t$'s.
It can be seen that the test accuracies of all algorithms decrease as $\gamma$ increases.
This is intuitive, as enlarged fluctuations impose new challenges.
It is observed that FedPBC outperforms all algorithms that are {\em not} aided by memory.

\noindent{\bf Impact of a cutting-off lower bound $\delta$.}
Recall that $p_i$'s might be too small and close to $0$ due to the unbalanced class contributions in $\bm{r}$.
We show in Lemma \ref{lemma: ergodicity} that a smaller lower bound $c$ of $p_i^t$'s slows down convergence and incurs a looser bound in~\prettyref{thm: main}.
Notice that FedPBC remains the best among the algorithms {\em not} aided by memory in terms of test accuracy.
At one challenging extreme (when $\delta = 0.001$),
all algorithms experience significant drops in accuracy,
in particular MIFA.
This confirms our conjecture that the old gradient might lead to staled updates and affect performance.

\noindent{\bf Impact of contribution heterogeneity $\sigma_0$.}
A smaller $\sigma_0$ leads to a more even contribution of each class and thus more homogeneous $p_i$'s.
Hence, 
it is not surprising to find that many baseline algorithms attain accurate test predictions when $\sigma_0 = 1.0$.
In contrast,
FedPBC shadows all baseline algorithms except MIFA 
in the highly heterogeneous scenario where $\sigma_0 = 20.0$.

\end{document}